\documentclass[journal]{IEEEtran}

\usepackage{booktabs}
\usepackage{amsmath}
\DeclareMathOperator*{\argmax}{arg\,max}
\usepackage{array}
\usepackage{graphicx}
\usepackage{bm} 
\usepackage{algorithmic}
\usepackage{multirow}
\usepackage{verbatim}
\usepackage[normalem]{ulem}
\useunder{\uline}{\ul}{}
\usepackage{stfloats}
\begin{document}

\title{SITUP: Scale Invariant Tracking using Average Peak-to-Correlation Energy}

\author{Haoyi Ma,~\IEEEmembership{Student Member,~IEEE,}
        Scott T. Acton,~\IEEEmembership{Fellow,~IEEE,}
        Zongli Lin,~\IEEEmembership{Fellow,~IEEE}

\thanks{Haoyi Ma, Scott T. Acton and Zongli Lin are with the Charles L. Brown
Department of Electrical and Computer Engineering, University of
Virginia, Charlottesville, VA 22904-4743, USA (e-mails: hm5au@virginia.edu, acton@virginia.edu and zl5y@virginia.edu).}
\thanks{\textbf{This work has been submitted to the IEEE for possible publication. Copyright may be transferred without notice, after which this version may no longer be accessible.}}}


\maketitle
\begin{abstract}

 Robust and accurate scale estimation of a target object is a challenging task in visual object tracking. Most existing tracking methods cannot accommodate large scale variation in complex image sequences and thus result in inferior performance. In this paper, we propose to incorporate a novel criterion called the average peak-to-correlation energy into the multi-resolution translation filter framework to obtain robust and accurate scale estimation. The resulting system is named SITUP: Scale Invariant Tracking using Average Peak-to-Correlation Energy. SITUP effectively tackles the problem of fixed template size in standard discriminative correlation filter based trackers. Extensive empirical evaluation on the publicly available tracking benchmark datasets demonstrates that the proposed scale searching framework meets the demands of scale variation challenges effectively while providing superior performance over other scale adaptive variants of standard discriminative correlation filter based trackers. Also, SITUP obtains favorable performance compared to state-of-the-art trackers for various scenarios while operating in real-time on a single CPU.

\end{abstract}


\IEEEpeerreviewmaketitle

\section{Introduction}

\IEEEPARstart{V}{isual} object tracking (VOT) is a well known problem in video analysis and computer vision with applications ranging from video surveillance and video compression to medical imaging \cite{video surveillance}\cite{video compression}\cite{ray2002tracking}\cite{cui2006monte}\cite{liu2012grid}\cite{mukherjee2004level}. Given the initial state of a target object in an initial frame, the goal of tracking is to estimate the state of the target in the following frames. Despite significant progress in recent years, the tracking problem is still not fully conquered as numerous complicated interfering factors affect the performance of a tracking algorithm, such as illumination variation, shape deformation, partial and full occlusion, to name a few. Since VOT is the basic building block of many time-critical systems, another major challenge is that a visual tracker should meet the strict constraints of time and computational budget, especially with respect to mobile operating systems or embedded computing architectures where real-time analysis is often desired and resources are limited.

VOT methods generally can be divided into two different groups: \textit{generative} and \textit{discriminative}. Generative tracking methods learn a model to represent the appearance of a target object, and then the tracking problem is formulated as finding the object appearance most similar to the model. Examples of generative tracking algorithms are found in \cite{IVT}, \cite{l1tracking} and \cite{gen1}. Instead of building a model to describe the appearance of an object, discriminative tracking methods aim to discriminate the target object from the background and are often found to outperform generative methods in accuracy. Discriminative trackers can usually run efficiently using inexpensive hand-crafted features \cite{HOG}\cite{color} and various learning methods such as a structured output support vector machine (SVM) \cite{STRUCK}, multi-expert entropy minimization \cite{MEEM}, and discriminative correlation filters \cite{MOSSE}\cite{KCF}\cite{CSK}.

Saliently, the discriminative correlation filter (DCF) is applied in a family of tracking methods characterized by both high accuracy and high efficiency. These methods approximate the dense sampling scheme by generating a circulant matrix, of which each row denotes a circular shift of the base sample. In such a case, the regression model can be computed in the Fourier domain, which brings significant speed improvement in both training and testing processes.

Since the DCF based tracker is template based and uses a fixed template size, the output states include only the vertical and horizontal locations of the target object in a video frame in the standard setting \cite{MOSSE}\cite{KCF}\cite{CSK}. However, in many applications, such as video surveillance and medical imaging, scale estimation of the target object is also important. Variations of the target object size may occur due to changes in the target object appearance or motion along the camera axis. A robust and accurate scale estimation is challenging and is further complicated by the presence of other challenging factors such as motion blur, partial and full occlusion.

\begin{figure*}[t]
\centering
\vspace{3pt}
\includegraphics[width=43mm,height=43mm]{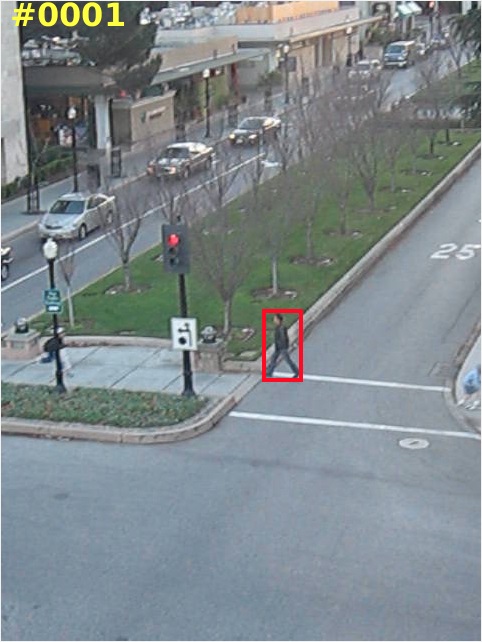}
\includegraphics[width=43mm,height=43mm]{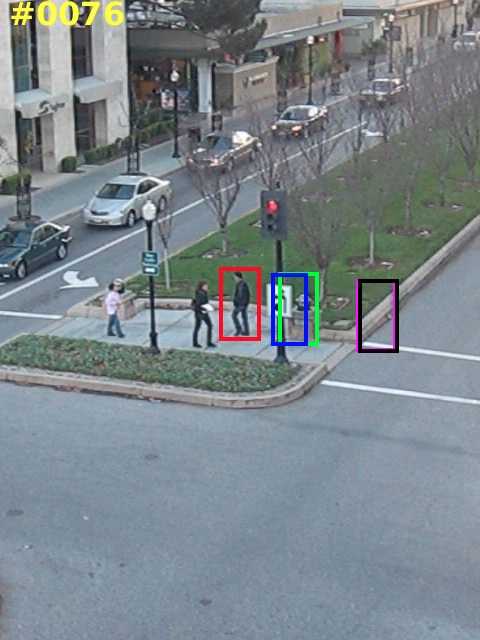}
\includegraphics[width=43mm,height=43mm]{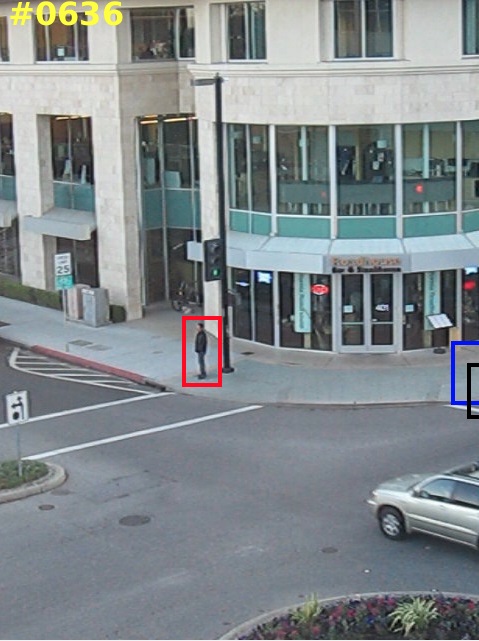}
\includegraphics[width=43mm,height=43mm]{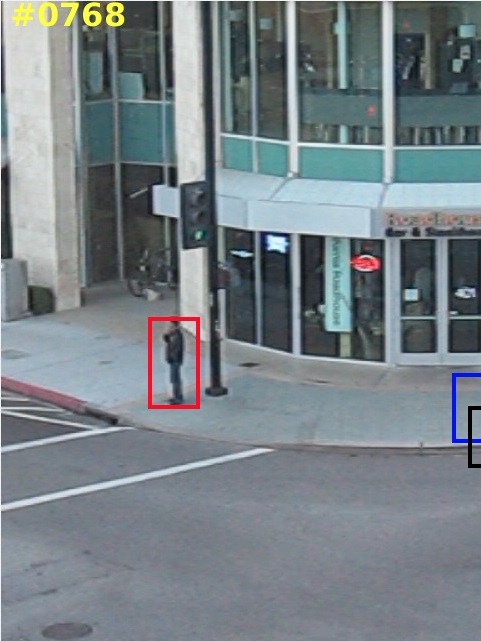}

\includegraphics[width=43mm,height=43mm]{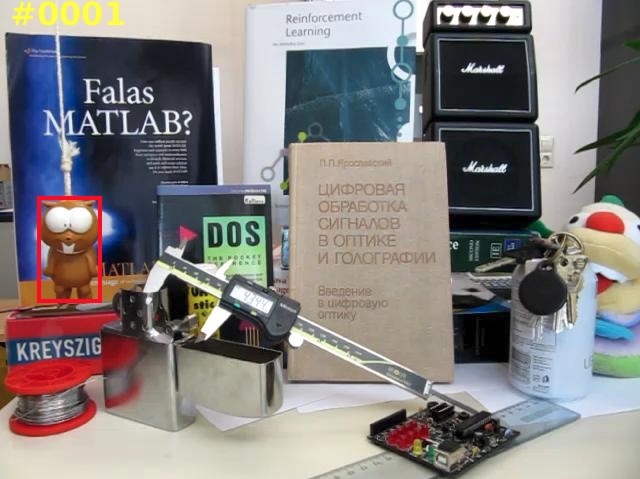}
\includegraphics[width=43mm,height=43mm]{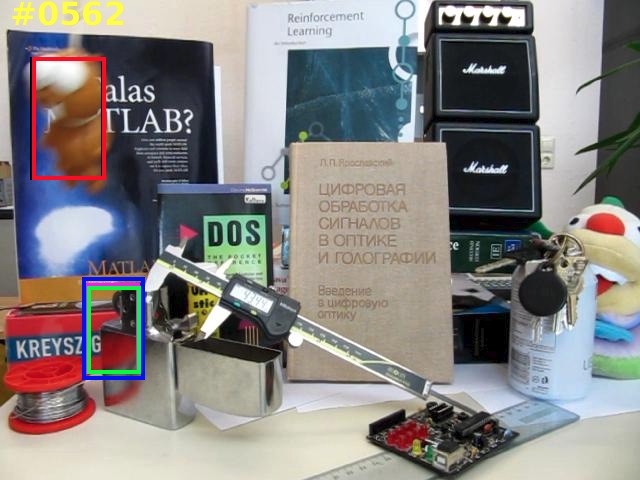}
\includegraphics[width=43mm,height=43mm]{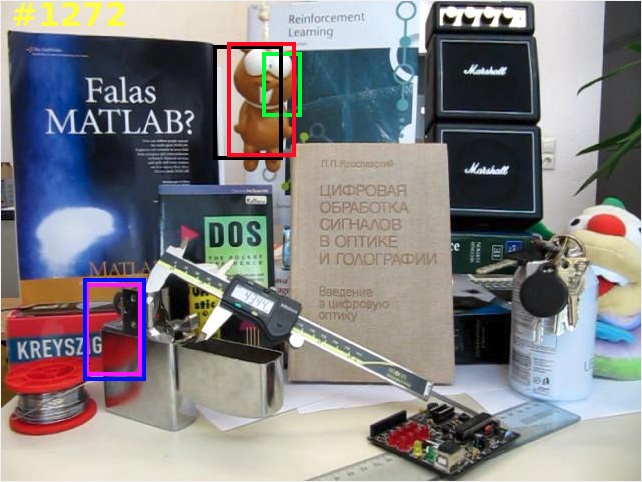}
\includegraphics[width=43mm,height=43mm]{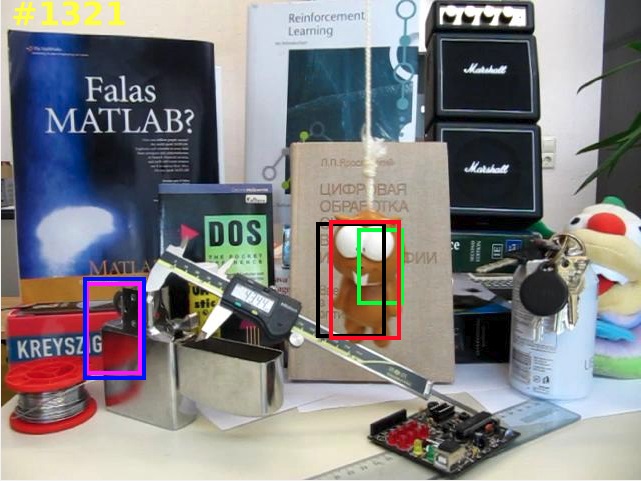}

\includegraphics[width=43mm,height=43mm]{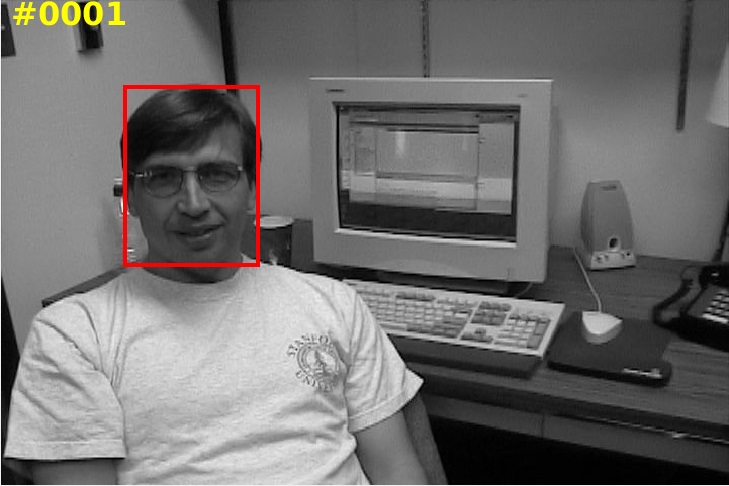}
\includegraphics[width=43mm,height=43mm]{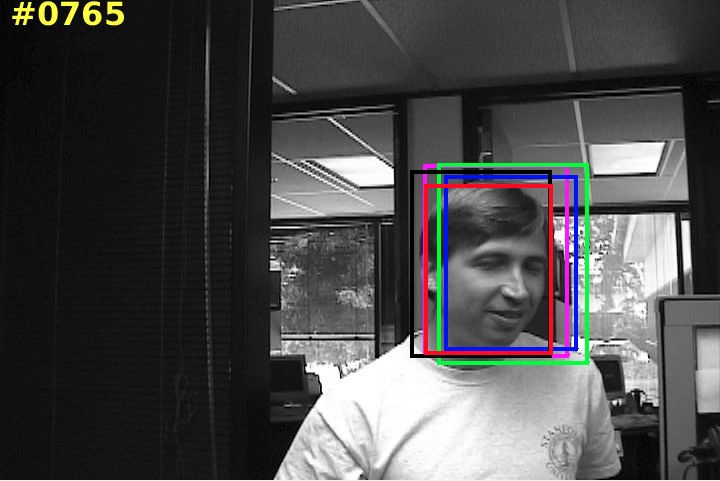}
\includegraphics[width=43mm,height=43mm]{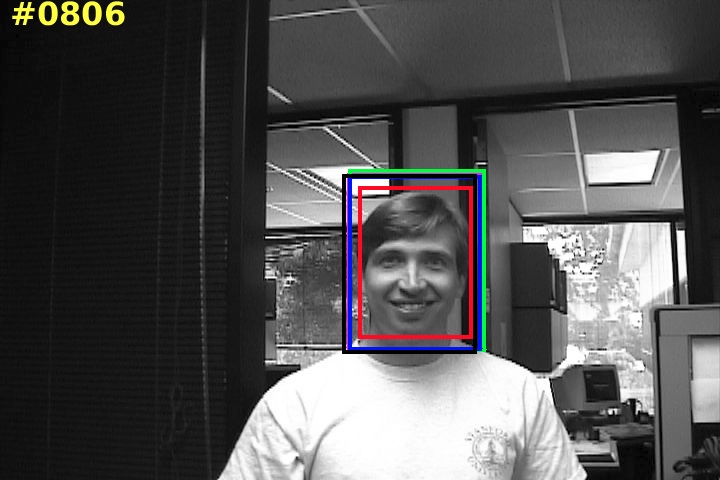}
\includegraphics[width=43mm,height=43mm]{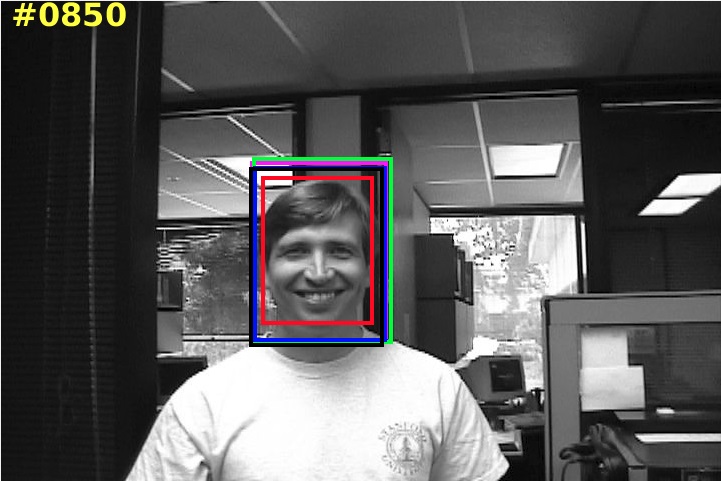}
\includegraphics[width=80mm,height=13mm]{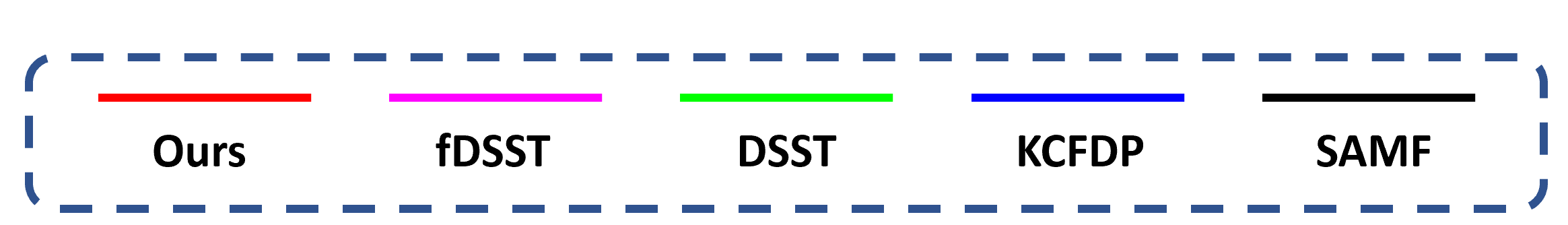}
\caption{Comparison of our Scale Invariant tracker Using average Peak-to-correlation energy (SITUP) with other scale adaptive variants of standard DCF based trackers, including scale adaptive tracker with multiple features (SAMF), discriminative scale space tracking (DSST), fast discriminative scale space tracking (fDSST) and kenerlized correlation filter with detection proposals (KCFDP). Example frames shown are from \textit{human3} (top row), \textit{lemming} (middle row) and \textit{dudek} (bottom row) sequences that are selected from online tracking benchmark datasets to involve the scale variation challenge. The bounding boxes in the first column identify the target objects to be tracked in corresponding sequences, and the number on the upper-left corner of each image is the frame number of corresponding image. Comparing to other scale adaptive variants of standard DCF based trackers, our approach significantly increases both the robustness and accuracy by estimating the target object size accurately. }
\label{Fig1}
\end{figure*}

To incorporate scale estimation into a standard DCF based tracker (to be described in Section \ref{The KCF tracker} and Section \ref{Multi-channel feature integration}), several approaches have been proposed \cite{DSST&fDSST}\cite{zhang2015robust}\cite{kcfdp}\cite{skcf}\cite{SAMF}\cite{DSST}. One straightforward approach is the scale adaptive tracker with multiple features (SAMF) \cite{SAMF}, which tackles the scale estimation problem by applying a standard two-dimensional correlation filter at multiple resolutions. Another approach is the joint scale spatial filter \cite{zhang2015robust}, which incorporates scale estimation by proposing a three-dimensional correlation filter in the joint spatial and scale space. This estimation is achieved by modeling the exhaustive template searching in the joint space as a block-circulant matrix. The discriminative scale space tracker (DSST) \cite{DSST} tackles the scale estimation problem by learning two separate correlation filters for explicit translation and scale estimation. Based on the DSST, the fast discriminative scale space tracker (fDSST) \cite{DSST&fDSST} further improves both accuracy and speed by way of feature dimension reduction, sub-grid interpolation of correlation scores and search space expansion. A kernelized correlation filter with detection proposals (KCFDP) is proposed in \cite{kcfdp} to tackle the aspect ratio variation problem in scale estimation by incorporating a class-agnostic detection proposal method into a standard DCF based tracker. An adjustable Gaussian window function and a keypoint based model are proposed in \cite{skcf} to tackle the scale estimation problem. Although they address the scale estimation problem to some degree, these methods suffer from the inferior performance compared to the state-of-the-art trackers.

In all the aforementioned scale estimation methods, the criterion used to obtain the scale estimation is the na\"ive maximum response value. In contrast, we argue that the robustness of the maximum response value will be heavily degraded due to the presence of some other challenging factors such as motion blur, partial and full occlusion. Here, in SITUP, instead of using the na\"ive maximum response value to obtain both the translation and scale estimation as in the aforementioned methods, we propose to use the \textit{average peak-to-correlation energy} (APCE) as the criterion to obtain the scale estimation. The exhaustive scale searching strategy is employed in our tracker. First, at the training stage, a two-dimensional correlation filter is trained in the standard DCF setting. At the testing stage, we obtain the sample patches centered around the estimated target location of the previous frame with different scales and resize them into a fixed size. The correlation filter response maps are obtained by comparing the resized sample patches with the learned model. The APCE measure of the response map at each scale is computed and the response map with the largest APCE measure is selected as the one with the best scale estimation. The translation estimation is obtained by searching for the position of the maximum response value within the response map with the largest APCE measure. After obtaining the translation and scale estimation, we update the learned model using the interpolation update strategy.

To validate the performance of our tracker, comprehensive quantitative and qualitative evaluations are performed on full online tracking benchmark (OTB) datasets: OTB2013, OTB50 and OTB100 \cite{OTB50}\cite{OTB100}. First, we show the effectiveness and robustness of our scale searching strategy by comparing our tracker SITUP with other scale adaptive variants of standard DCF based trackers. Fig. 1 shows a qualitative comparison of our approach with other scale adaptive variants of standard DCF based trackers. It can be seen that our approach accurately estimates the target size and thereby significantly improves the robustness and accuracy. Further, we compare our tracker with 10 state-of-the-art trackers and analyze both the overall and attribute-based performance to demonstrate the robustness of our tracker when dealing with different scenarios. It is worth mentioning that our proposed scale estimation scheme is generic and can be incorporated into any DCF based tracker. For future development, the results and Matlab code will be available to the public at \textit{https://github.com/haoyihaoyi/Scale-adaptive-tracker}.

The rest of this paper is organized as follows. Section \ref{related work} gives an overview of the prior works most relevant to our proposed approach. In Section \ref{The KCF tracker} and Section \ref{Multi-channel feature integration}, we introduce the multi-channel kernelized correlation filter, which we adopt as the baseline tracker. Our scale searching strategy is described in Section \ref{SITUP}. In Section \ref{111}, the implementation details are presented so that our results can be reproduced. The benchmark datasets and evaluation protocols are described in Section \ref{222}. The comparison results of the performed experiments are presented in Section \ref{333} and Section \ref{444}. Finally, Section \ref{conclusion} concludes the paper.

\section{Related Work}
\label{related work}
VOT is one of the fundamental problems in image and video processing. Here, we consider the single object tracking task where the target object is identified in the first video frame and is to be tracked in the following frames. This problem is challenging since the target object is class-agnostic and is defined solely by its initial location and scale.

Typically, VOT methods work by building a target appearance model from the observed image information by way of a generative \cite{IVT}\cite{l1tracking}\cite{gen2} or a discriminative model \cite{MOSSE}\cite{KCF}\cite{CSK}. Generative appearance models describe the target appearance with statistical models or templates. Discriminative tracking methods instead implement machine learning techniques to discriminate the target appearance from the surrounding background by formulating VOT as a classification problem.

DCF based methods have been successfully applied to visual tracking \cite{MOSSE}\cite{KCF}\cite{CSK} and have achieved high efficiency and high accuracy. These methods have shown to provide excellent results on tracking benchmark datasets \cite{OTB50}\cite{OTB100}\cite{VOT2014}, while maintaining real-time speed. The DCF based trackers locate the target object in the new frame by learning a discriminative correlation filter to discriminate the target object from the background. Bolme \textit{et al}. \cite{MOSSE} proposed to train the correlation filter via minimizing the total squared error between the actual and the desired correlation outputs on a set of grayscale sample patches. By utilizing circular correlation, the authors showed that the resulting correlation can be computed efficiently in the Fourier domain, which brings hundreds of time speed improvement in both the training and testing stages. Henriques \textit{et al}. \cite{CSK} \cite{KCF} further showed that the DCF framework can be equivalently formulated as a regression problem on the set of all cyclic shifts of the training sample patches.

To generalize the DCF framework to include multi-dimensional features, several solutions have been proposed that attempt to learn an exact multi-channel filter from the set of the training samples. However, computational cost has plagued such approaches. To remediate, approximate formulations for learning multi-channel filters \cite{KCF}\cite{CN} have been proposed for visual object tracking that have proven to be robust while scaling linearly with the number of channels.

In many different tracking scenarios, the DCF based approaches have demonstrated the capability of locating the target accurately. However, the standard DCF based tracker is restricted to translation estimation since it is a template based method with a fixed template size. The standard DCF based tracker will yield inferior performance when a large scale variation of the target object is encountered. Hence, achieving accurate scale estimation of the target object is beneficial to tracking in many aspects by providing a more accurate tracking result.

Recently, Danelljan \textit{et al}. \cite{DSST} proposed DSST method that, by directly learning the appearance changes induced by scale variation, trains two separate correlation filters for translation estimation and scale estimation respectively. A two-dimensional standard discriminative correlation filter was employed for translation estimation and another one-dimensional discriminative correlation filter was employed for scale estimation. Further, an improved version fDSST was proposed in \cite{DSST&fDSST} to increase the speed and accuracy of DSST. To reduce the computational cost, the principal component analysis (PCA) was employed to reduce the feature dimension, leading to a reduction of the required number of fast Fourier transform (FFT). Sub-grid interpolation was employed to allow the use of coarser feature grids for both the training and the detection stages, resulting in the reduction of the size of the performed FFTs required to reduce the computational cost. The significant speed improvement provided the flexibility to improve the robustness by expanding the search space of the translation filter. However, the unreliable translation estimation might lead to an inferior scale estimation in both DSST and fDSST, which results in a poor generalization capability. Alternatively, Li and Zhu proposed the SAMF method \cite{SAMF} to tackle the scale estimation problem by extending a standard DCF based tracker for translation estimation to multiple resolutions, and the scale and translation estimation are obtained jointly. SAMF is incorporated into many DCF based tracking methods due to its good generalization capability. To deal with the aspect ratio variation problem in scale estimation, Huang \textit{et al}. \cite{kcfdp} proposed the KCFDP method to incorporate a class-agnostic detection proposal method into the standard discriminative correlation filter based tracker. EdgeBoxes \cite{edgebox} is employed to enable the scale and aspect ratio adaptability of a standard translation correlation filter with the assumption that the number of contours that are wholly contained in a bounding box is indicative of the \textit{objectness}. The proposal generator EdgeBoxes \cite{edgebox} is tuned and coupled with the proposal rejection strategy to provide promising candidates with different scales and aspect ratios. However, both SAMF and KCFDP suffer from inferior performance compared to the state-of-the-art trackers.

SAMF is the solution most closely related to our proposed approach, since we also employ a multi-resolution scale searching strategy. Here, instead of using the na\"ive maximum response value as in \cite{SAMF}, we propose to employ the APCE measure as the criterion to obtain the best scale estimation first and the translation estimation is further obtained. Due to the robustness of the APCE measure when facing different scenarios, our approach obtains superior performance while maintaining a real-time speed despite its utter simplicity.

\section{Scale Invariant Tracking using Average Peak-to-Correlation Energy}
\label{our approach}
In Section \ref{The KCF tracker} and Section \ref{Multi-channel feature integration}, the formulation of our baseline tracker multi-channel kernelized correlation filter is treated. Our scale searching strategy is described in detail in Section \ref{SITUP}. 

\subsection{The kernelized correlation filter tracker}
\label{The KCF tracker}
SITUP leverages the kernelized correlation filter (KCF) tracker \cite{KCF}. By exploiting the structure of the circulant matrix with high efficiency, the discriminative ability of the KCF tracker is enhanced with the augmentation of negative samples while maintaining high speed. 

In the formulation of KCF, the generated data matrix has a circular structure by using the cyclic shifts of the base sample to approximate the dense samples over the base sample. For notational simplicity, we start with one-dimensional signals. Given one dimensional data $\mathbf{x} = [\mathbf{x}_1, \mathbf{x}_2, ..., \mathbf{x}_n]$. A cyclic shift of $\mathbf{x}$ will be $P\mathbf{x} = [\mathbf{x}_n, \mathbf{x}_1, \mathbf{x}_2, ..., \mathbf{x}_{n-1}]$, where $P$ is a permutation matrix. The full set of the cyclic shifts of base sample can be written as $\{P^{u}\mathbf{x}| u = 0, 1, ..., n-1\}$, which can further be concatenated to generate the data matrix $X = C(\mathbf{x})$. Matrix $C(\mathbf{x})$ is called a circulant matrix since the matrix is purely generated by the cyclic shifts of $\mathbf{x}$. One intriguing property of the circulant matrix is that all circulant matrices can be diagonalized by the discrete Fourier transform (DFT) \cite{gray2006toeplitz}, regardless of the generating vector $\mathbf{x}$. This can be expressed as
\begin{equation}
    X = F {\rm{diag}}{(\hat{\mathbf{x}}) F^{H}}, 
\end{equation}
where $F$ is the DFT matrix that is independent of $\mathbf{x}$, $F^{H}$ is the Hermitian transpose of $F$ and $\hat{\mathbf{x}}$ denotes the DFT of the generating vector, $\hat{\mathbf{x}} = F(\mathbf{x})$.

In KCF, the goal of training is to find $f(\mathbf{z}) = \mathbf{w}^{T}\mathbf{z}$ that minimizes the squared error over samples $\mathbf{x}_i$ and their regression targets $\mathbf{y}_i$ as follows:

\begin{equation}
    \min_{\mathbf{w}} \sum_{i}^{n} (f(\mathbf{x}_i)-\mathbf{y}_i)^2 + \lambda \|\mathbf{w}\|. 
\end{equation}
The scalar $\lambda$ is a regularization parameter that controls overfitting. This ridge regression problem has a closed-form solution, which can be written as

\begin{equation}
\mathbf{w} = (X^{H}X + \lambda I)^{-1}{X^H}\mathbf{y}.   
\end{equation}
Each row of the data matrix $X$ corresponds to one sample $\mathbf{x}_i$, each element of $\mathbf{y}$ is a regression target $\mathbf{y}_i$, and $I$ is an identity matrix. The decomposition of the circulant matrix can be utilized to simplify the solution of the regression problem. Substitution of (1) in (3) results in the solution,
\begin{equation}
    \hat{\mathbf{w}}^{*} = \frac{\hat{\mathbf{x}}^{*}\odot\hat{\mathbf{y}}}{\hat{\mathbf{x}}^{*}\odot\hat{\mathbf{x}} + \lambda},
\end{equation}
where ${\hat{\mathbf{x}}}^{*}$ is the complex-conjugate of $\hat{\mathbf{x}}$. This solution reduces the computational cost of both extracting patches explicitly and solving a general regression problem, since it only involves the DFT and element-wise operation. 

To allow for a more powerful classifier with nonlinear regression functions $f(\mathbf{z})$, the solution $\mathbf{w}$ is expressed as
\begin{equation}
    \mathbf{w} = \sum_{i} \alpha_i\varphi(\mathbf{x}_i),
\end{equation}
where $\alpha_i$'s are the variables under optimization in dual space, as opposed to the primal space $\mathbf{w}$.
The kernel trick can be employed as
\begin{equation}
    f(\mathbf{z}) = \mathbf{w}^{T}\mathbf{z} = \sum_{i=1}^{n}{\alpha_i}\kappa(\mathbf{z},\mathbf{x}_i).
\end{equation}
For the most commonly used kernels (e.g., Gaussian, linear and polynomial), the circulant matrix trick can also be employed, since these kernels treat each dimension of the data equally \cite{KCF}. The dual space coefficients $\bm{\alpha} = [\alpha_1, \alpha_2, ..., \alpha_n]$ can be computed as
\begin{equation}
    {\hat{\bm{\alpha}}}^{*} = \frac{\hat{\mathbf{y}}}{\hat{\mathbf{k}}^{\mathbf{x}\mathbf{x}}+\lambda},
\end{equation}
where $\mathbf{k}^{\mathbf{x}\mathbf{x}}$ is the kernel correlation as defined in \cite{KCF}. In SITUP, the linear kernel is adopted in consideration of both speed and accuracy as
\begin{equation}
    \mathbf{k}^{\mathbf{x}{\mathbf{x}^{'}}} = \mathcal{F}^{-1}(\hat{\mathbf{x}}^{*}\odot\hat{\mathbf{x}}^{'}).
\end{equation}

\begin{figure*}[t]
\centering
\includegraphics[width=160mm,height=70mm]{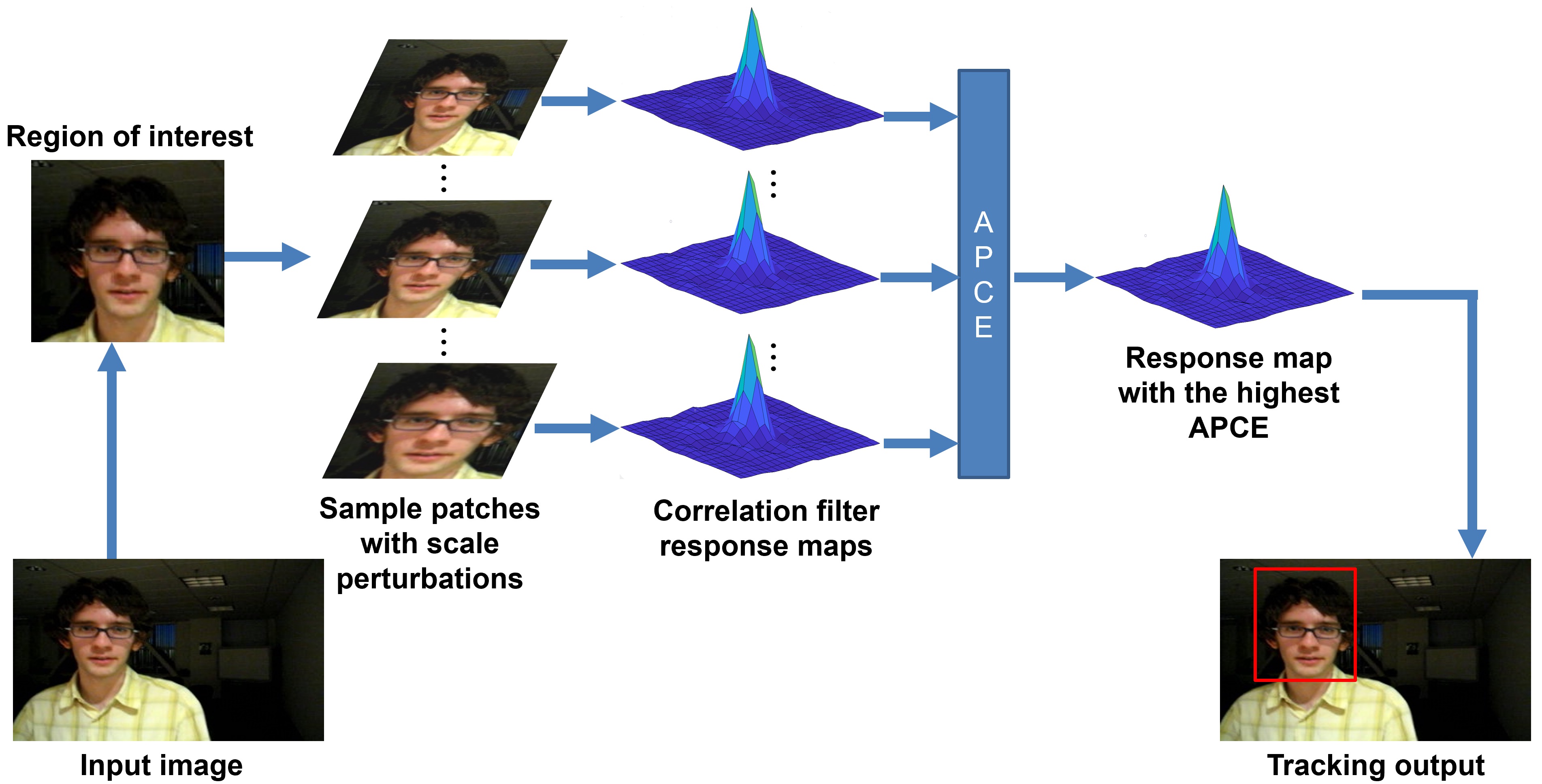}
\caption{Main steps of the proposed algorithm. Patches with different scales are sampled centered around the previous estimated target object location and registered to the same size as the fixed size. The response maps are computed respectively and the response map with the highest APCE measure is chosen as the one with the best scale estimation. The translation estimation is obtained by localizing the maximum response value within the response map with the highest APCE measure.}
\label{Fig2}
\end{figure*}

Since the algorithm only requires element-wise operation and DFT/IDFT, the computational cost is at a nearly-linear $\mathcal{O}(n\log n)$. A large search window is employed to enclose more negative samples for training. Besides training, the circulant matrix trick can also be employed to speed up the detection process. In the next frame, the patch $\mathbf{z}$ at the same location is treated as the base sample and the corresponding response map in the Fourier domain is computed as
\begin{equation}
    \hat{\mathbf{f}}(\mathbf{z}) = \hat{\mathbf{k}}^{\mathbf{x}\mathbf{z}}\odot\hat{\bm{\alpha}},     
\end{equation}
where $\mathbf{x}$ and $\mathbf{z}$ correspond to the base sample and the base patch respectively. The response map in the spatial domain is computed through IDFT as
\begin{equation}
    {\mathbf{f}}(\mathbf{z}) = \mathcal{F}^{-1} (\hat{\mathbf{f}}(\mathbf{z})).
\end{equation}
When using IDFT to transform $\hat{\mathbf{f}}(\mathbf{z})$ back to the spatial domain, the translation with respect to the maximum response is treated as the translation of the target object. Since the template is of fixed size, the update process is straightforward. Two sets of coefficients should be updated, one being the dual space coefficients $\bm{\alpha}$, and the other the base sample $\mathbf{x}$. As the setting in \cite{KCF}, the coefficients are updated through linear interpolation using
\begin{equation}
    \mathbf{T} = \theta \mathbf{T}_{\rm{new}} + (1 - \theta)\mathbf{T}, 
\end{equation}
where $\mathbf{T} = [\bm{\alpha}^{{T}}, \mathbf{x}^{T}]^{T}$ is the template to be updated.

\subsection{Multi-channel feature integration}
\label{Multi-channel feature integration}

Working in the dual space $\bm{\alpha}$ instead of the primal space $\mathbf{w}$ has the advantage of allowing multi-channel features. The kernel correlation function only needs to compute norms of the arguments. By summing the individual dot-product for each channel in the Fourier domain, a dot-product can be computed easily. Suppose that the multiple channels of the data representation are concatenated into a vector as $\mathbf{x} = [\mathbf{x}_1, \mathbf{x}_2 ..., \mathbf{x}_{c}]$, the multi-channel extension of the linear kernel can be rewritten as
\begin{equation}
    \mathbf{k}^{\mathbf{x}\mathbf{x}^{'}} = \mathcal{F}^{-1}\left(\sum_{c}\hat{\mathbf{x}}^{*}_{c}\odot\hat{\mathbf{x}}^{'}_{c}\right).
\end{equation}
Strong features rather than the grayscale pixel values can be employed in the tracker. Three types of features are utilized in the SITUP tracker including raw grayscale pixel values, histogram of oriented gradients, and color-naming.

The histogram of oriented gradients (HoG) \cite{HOG} is effective for object detection and can be computed efficiently. The image is divided into small connected regions named cells and the gradient information of uniformly spaced cells is extracted. Then, the HoG counts occurrences of discrete gradient orientation to form the histogram.

Color names (CN) or color-naming are linguistic color labels, which are assigned by human to describe colors in the world. Recently, CNs have been adopted in different vision tasks such as action recognition \cite{khan2013coloring} and visual object tracking \cite{CSK}, and have achieved promising results. A CN provides the perception of the target object color, which contains salient information regarding the target object as shown in \cite{staple} and \cite{siena2016detecting}.

\subsection{Our scale searching framework}
\label{SITUP}
Our scale searching scheme bears some similarity to the structure of SAMF, since we also employ a multi-resolution translation filter framework. In contrast to the SAMF approach, SITUP exploits the robustness of APCE criterion, which enables robust and accurate scale estimation when facing different scenarios. SAMF only employs a na\"ive maximum response value as the criterion, with which the robustness will be heavily degraded when scale variation presents with other challenging factors. Besides, our approach obtains a superior performance with a reliable update of the target appearance model since the APCE measure can reflect the confidence level of the response map.

The fluctuation and the peak value of the response map can reveal the confidence level of the tracking result. In the extreme case, the ideal response map should have only one sharp peak at the true target object location while being smooth and close to zero in all other areas when the detected target object is perfectly matched to the correct target object location and scale. A sharper peak will result in a higher localization accuracy. On the contrary, the response map will fluctuate fiercely, and the APCE measure will significantly decrease when challenged with severe occlusion and motion blur. Wang \textit{et al}. \cite{LMCF} proposed to employ the maximum response value and an APCE measure to provide a high-confidence update strategy for robustness. The APCE measure is defined as
\begin{equation}
    APCE = \frac{{|\mathbf{f}_{\max}-\mathbf{f}_{\min}|}^2}{{\rm{mean}}(\sum_{w,h}(\mathbf{f}_{w,h}-\mathbf{f}_{\min})^2)},
\end{equation}
where $\mathbf{f}_{\max}$, $\mathbf{f}_{\min}$, and $\mathbf{f}_{w,h}$ denote the maximum, the minimum and the $w$-th row $h$-th column elements of the response map $\mathbf{f}$ respectively. The APCE criterion is more robust than the maximum response value criterion, which is noticeable when scale variation and other challenges such as motion blur and occlusion occur at the same time. See Fig. 3, where example frames are from the \textit{BlurOwl} sequence, which involves the four challenges of scale variation, motion blur, fast motion and in-plane rotation. SITUP obtains accurate scale and translation estimation with the robust APCE measure. However, a tracker that employs the maximum response value as the criterion within the same framework cannot handle the scale variation due to the presence of other challenging factors, and thus yields poor performance.

\begin{figure}[t]
\centering
\includegraphics[width=28mm,height=25mm]{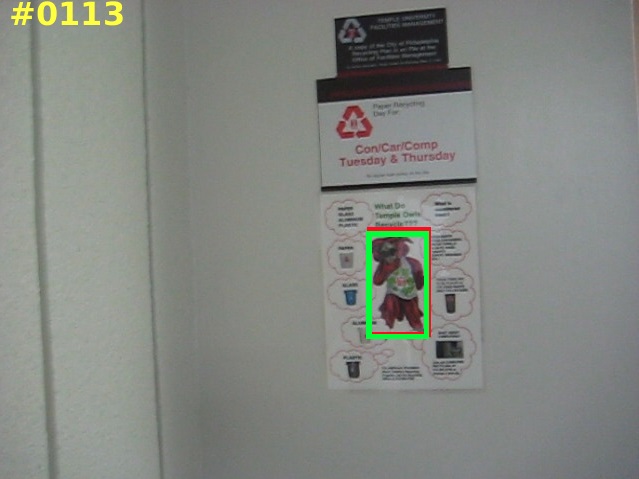}
\includegraphics[width=28mm,height=25mm]{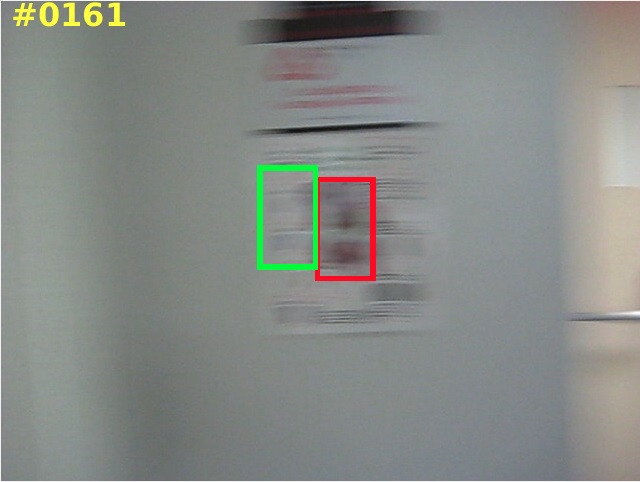}
\includegraphics[width=28mm,height=25mm]{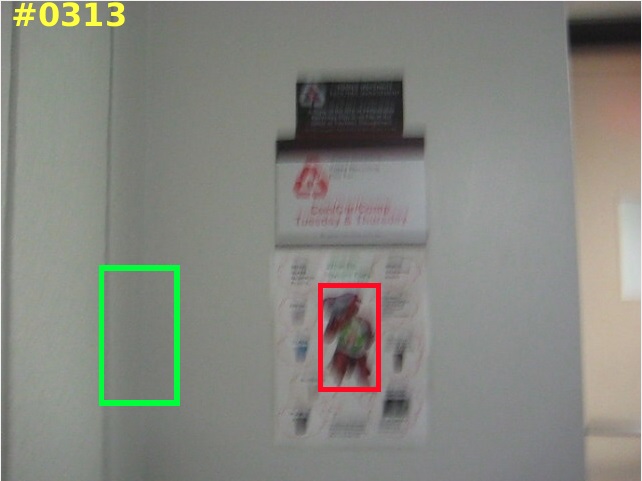}
\caption{Comparison of our tracker (in red) with the tracker which employs the maximum response value as the criterion within the same framework (in green). Benefiting from the robustness of the APCE measure, SITUP can accurately estimate the target size and location. Example frames shown are from the \textit{BlurOwl} sequence, which involves four challenges with regard to scale variation, motion blur, fast motion and in-plane rotation.}
\label{Fig3}
\end{figure}

As shown in Fig. 2, a standard two-dimensional discriminative correlation filter is trained first to provide preliminary translation estimation. Then, several patches centered around the estimated target location with different scale perturbations are sampled. All extracted patches are registered to the fixed template size. The response map of each patch is computed individually with the translation filter. The APCE measure of each response map is computed and the response map with the highest APCE measure is chosen as the one with the best scale estimation. The translation estimation is obtained by localizing the maximum response value within the response map with the highest APCE measure. With the new translation and scale estimation, the corresponding patch is extracted and employed to update the coefficient matrix and the target appearance model.

The template size is fixed as $\mathbf{s}_\mathbf{T} = (s_x, s_y)$. The scaling pool is defined as $\mathbf{S} = \{t_1, t_2, ..., t_k\}$. Assume that the target searching window size of the last frame is $\mathbf{s}_{{j-1}}$. For the current $j$th frame, $k$ patches with their sizes in $\{t_i\mathbf{s}_{j-1}|t_i \in \mathbf{S}\}$ are sampled and resized through bilinear-interpolation to the fixed template size $\mathbf{s}_\mathbf{T}$. The response map of a resized patch is defined as $\mathbf{f}_{t_i}$ and computed using (9) and (10). The corresponding APCE measure is defined as $APCE_{t_i}$ and is computed by (13). The scale estimation is obtained by searching for the response map with the highest APCE measure as 
\begin{equation}
    t = \argmax_{t_i}(APCE_{t_i}).
\end{equation}
The translation estimation is obtained by searching for the location of the highest response value within the response map $\mathbf{f}_{t}$. Since the target movement is implied in the response map, the final displacement should be tuned by $t$ to obtain the real translation bias. The update procedure will be implemented following (11). The main steps of SITUP are summarized in Algorithm 1.\\

\hrule height 0.8pt
\vspace{1.5pt}
\noindent {\textbf{Algorithm 1} Iteration on the $j$th frame in SITUP.}
\vspace{1.5pt}
\hrule
\begin{algorithmic}[1]
\REQUIRE ~~\\ 
Image $I_j$;\\
Previous target position, $p_{j-1}$;\\
Previous target searching window size $s_{j-1}$;\\
The template for the tracked target, $\mathbf{x}$;\\ 
The dual space coefficient, $\bm{\alpha}$.\\
\ENSURE ~~\\ 
The new target position, $p_j$;\\
The new target searching window size $s_j$;\\
The updated template for the tracked target, $\mathbf{x}$;\\
The updated dual space coefficient, $\bm{\alpha}$.\\
\FOR{every $t_i$ in $\mathbf{S}$} 
\STATE Sample the new patch $\mathbf{z}^{t_i}$ centered in $p_{j-1}$ with size $t_i\mathbf{s}_{j-1}$ and resize it to $\mathbf{s}_{\mathbf{T}}$.
\STATE Compute the response map $\mathbf{f}_{t_i}$ using (9) and (10).
\STATE Calculate the APCE measure $APCE_{t_i}$ using (13).
\ENDFOR
\STATE Obtain the new position $p_{j}$ and searching window size $s_{j}$ according to (14).
\STATE Compute $\mathbf{x}_{\rm{new}}$ based on the new position $p_{j}$ and searching window size $t_is_{j-1}$, and calculate $\bm{\alpha}_{\rm{new}}$ with (7).
\STATE Update $\mathbf{x}$ and $\bm{\alpha}$ with $\mathbf{x}_{\rm{new}}$ and $\bm{\alpha}_{\rm{new}}$.
\RETURN updated $\mathbf{x}$, $\bm{\alpha}$ with $p_j$ and $s_j$.
\end{algorithmic} 
\hrule

\section{Experiment}
\label{experiment}
Extensive evaluation is performed on three benchmark datasets to validate our approach. The implementation details are presented in Section \ref{111}. The benchmark datasets and evaluation protocols are described in Section \ref{222}. A comparison between our approach and other scale adaptive variants of standard DCF based trackers are presented in Section \ref{333}. To demonstrate contributions in performance, SITUP is compared to 10 state-of-the-art trackers in Section \ref{444}.

\begin{figure*}[t]
\centering
\includegraphics[width=55mm,height=45mm]{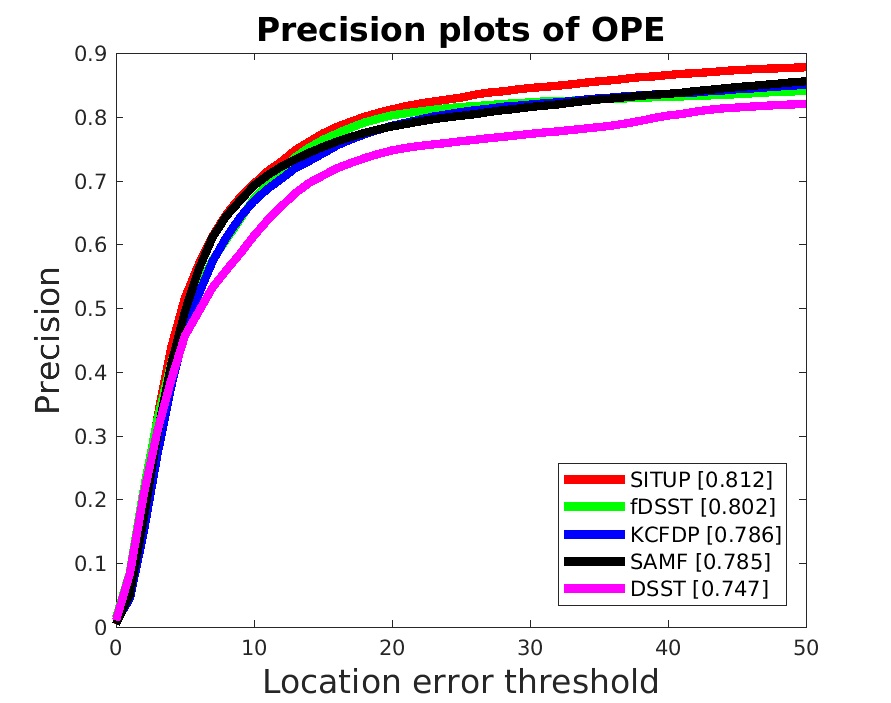}
\includegraphics[width=55mm,height=45mm]{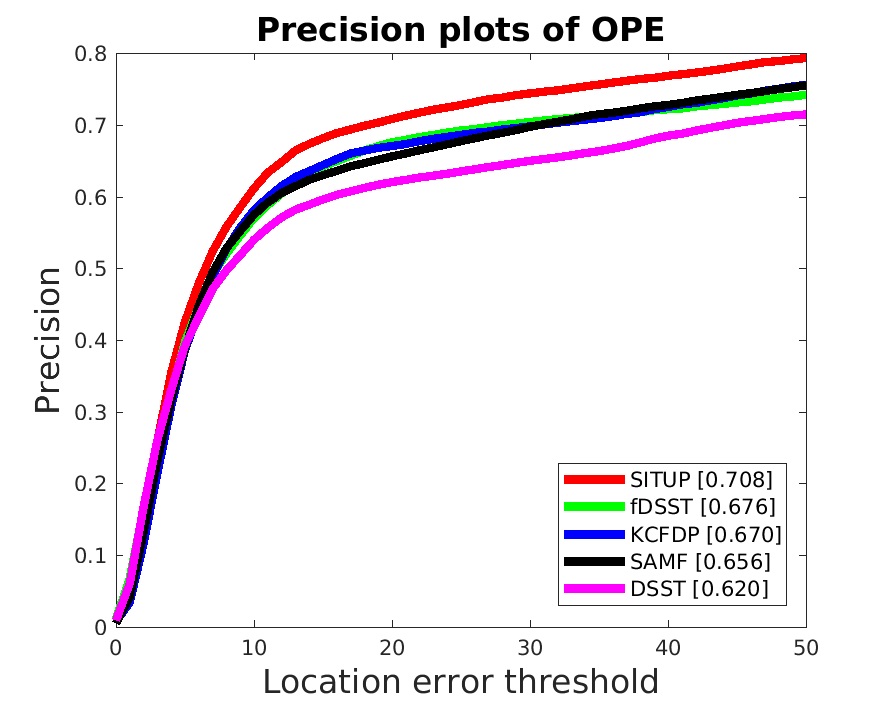}
\includegraphics[width=55mm,height=45mm]{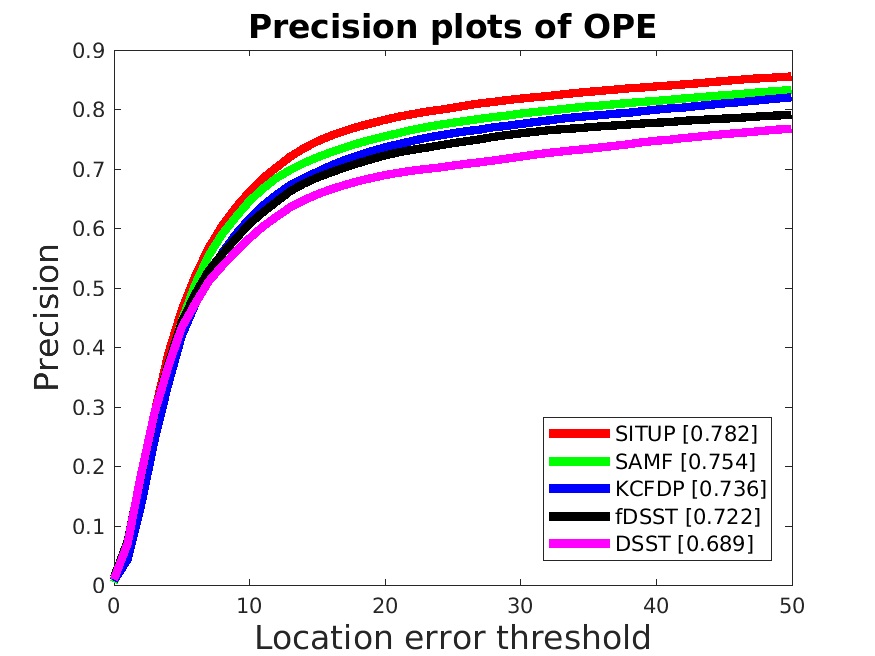}
\includegraphics[width=55mm,height=45mm]{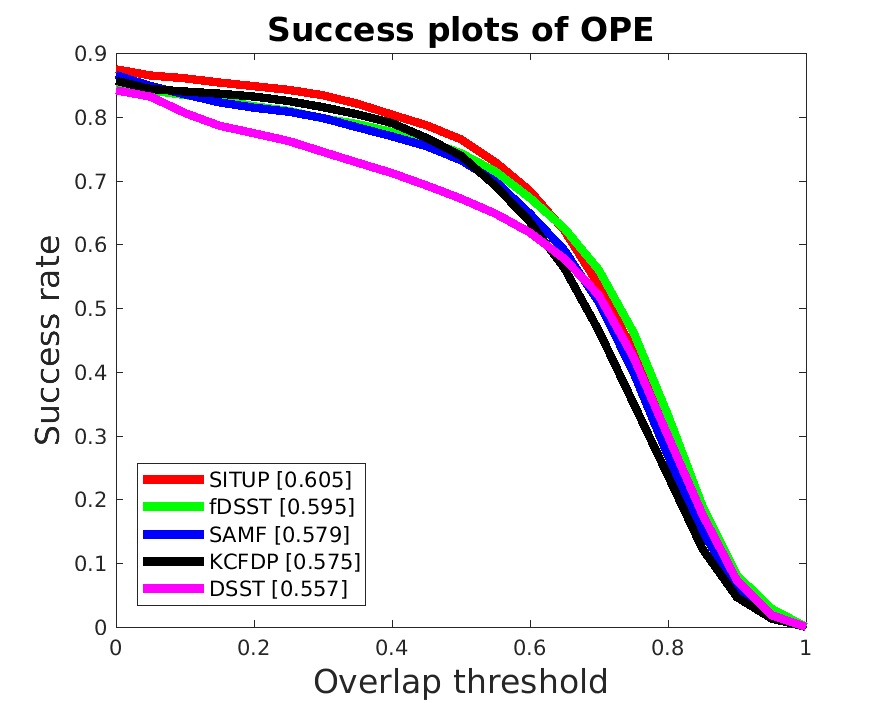}
\includegraphics[width=55mm,height=45mm]{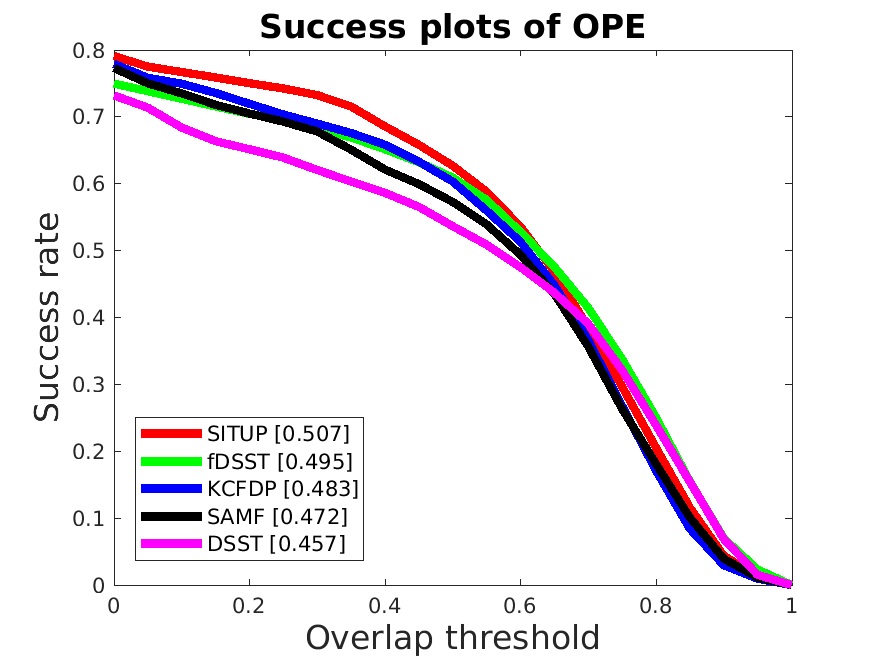}
\includegraphics[width=55mm,height=45mm]{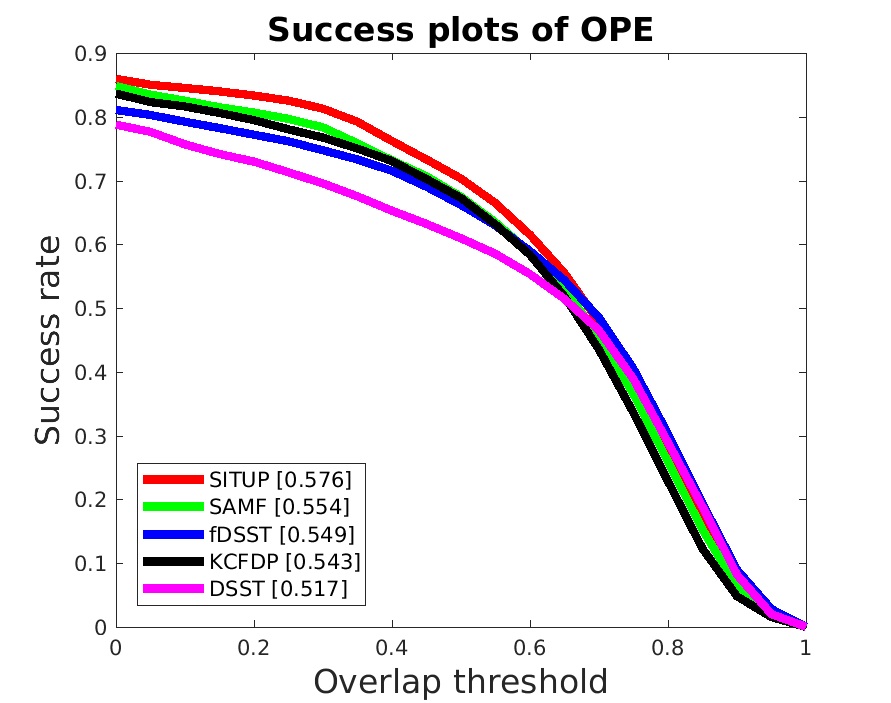}
\caption{The precision plots and success plots of our tracker SITUP and other scale adaptive variants of standard DCF based trackers. From left to right, the plots correspond to OTB2013, OTB50, OTB100. The top row are the precision plots and the bottom row are the success plots.}
\label{Fig4}
\end{figure*}

\begin{table}[t]
\caption{\upshape{Comparisons with other scale adaptive variants of standard DCF based trackers on OTB2013, OTB50 and OTB100 benchmark sequences. SITUP achieves the best performance in terms of the area-under-curve (AUC) score and the precision score at a standard threshold used in the literature of 20 pixels. The first and the second best values are highlighted in bold and underlined.}}
\label{table}
\begin{tabular}{cccccccc}
\hline
  &  & \multicolumn{2}{c}{\textbf{OTB2013}} & \multicolumn{2}{c}{\textbf{OTB50}} & \multicolumn{2}{c}{\textbf{OTB100}} \\
\textbf{Method} &\textbf{Speed}   & AUC             & Prec.  & AUC     & Prec.    & AUC     & Prec.\\ \hline
DSST            &     23.43        & 0.557            & 0.747   & 0.457   & 0.620    & 0.517   & 0.689             \\
fDSST           &\textbf{85.30} & \underline{0.595} & \underline{0.802} & \underline{0.495}   & \underline{0.676} & 0.549 & 0.722 \\
KCFDP           &30.27            & 0.575             & 0.786            & 0.483    & 0.670    & 0.543     & 0.736              \\
SAMF            &27.19            & 0.579             & 0.785            & 0.472    & 0.656    & \underline{0.554} & \underline{0.754}  \\
SITUP          &\underline{32.35} & \textbf{0.605} & \textbf{0.812} & \textbf{0.507} & \textbf{0.708} & \textbf{0.576} & \textbf{0.782}\\ \hline
\end{tabular}
\end{table}

\subsection{Implementation details}
\label{111}

The regularization parameter is set as $\lambda = 0.0003$, and the learning rate is set as $\theta = 0.004$. The standard deviation of the desired correlation output is set to be $1/10$ of the target size. The padding ratio is set as $1.5$ to make the search window $2.5$ times the target size in terms of both width and height, which is typical in the existing literature.

For the scaling pool, we employ the same setting $\mathbf{S} = \{0.985, 0.99, 0.995, 1.0, 1.005, 1.01, 1.015\}$ as in SAMF \cite{SAMF}. The features we employ are computed by augmenting HoG, CN, and grayscale pixel values. To reduce computational cost, a 31-dimensional variant of HoG is employed as in \cite{SAMF} and \cite{KCF}. The RGB space is transformed into a color-naming space, which is an 11-dimensional color representation. The grayscale features are normalized to the range $[-\frac{1}{2}, \frac{1}{2}]$. Each extracted feature channel of the sample patch is weighted by a cosine window to address the boundary effect. Also, we employ the linear kernel in consideration of both speed and accuracy in our tracker.

All tracking algorithms are implemented with Matlab on an Intel Core i7-7700K 4.5 GHz CPU with 16 GB RAM. A GeForce GTX 1080 GPU and MatConvNet 1.0-beta25 are used to reproduce the results of the trackers that require GPU implementation. Parameters of our tracking algorithm are fixed in all experiments. The parameters of other trackers are set according to the values in their original code.

\subsection{Datasets and metrics}
\label{222}
All trackers are quantitatively evaluated on the online tracking benchmark (OTB) datasets OTB2013, OTB50 and OTB100, following the one-pass evaluation (OPE) protocol described in \cite{OTB50} and \cite{OTB100}. OTB2013 is proposed in \cite{OTB50} for a comprehensive performance evaluation with 50 video sequences that is fully annotated with ground truth bounding boxes. To further address the problem of insufficient data, OTB100 is proposed in \cite{OTB100} with 100 fully annotated video sequences that contains previous 50 sequences proposed in \cite{OTB50}. Since some target objects are similar or less challenging, 50 difficult and representative ones of OTB100 are selected and represented as OTB50 for an in-depth analysis. The sequences in OTB datasets are categorized by way of 11 attributes. The 11 attributes in the OTB datasets are: illumination variation (IV), out-of-plane rotation (OPR), scale variation (SV), occlusion (OCC), deformation (DEF), motion blur (MB), fast motion (FM), in-plane rotation (IPR), out-of-view (OV), background clutter (BC), and low resolution (LR).

To quantitatively evaluate the tracking results on OTB datasets, three standard evaluation metrics are exploited including precision, success rate and tracking speed in frames per second (fps). The precision is the relative number of frames in a video sequence for which the estimated location is within the specified threshold distance of the ground truth. The precision plot can be obtained by evaluating the precision at different defined thresholds. The representative precision score for each tracker is reported at the threshold of 20 pixels as the setting in the standard OTB toolkit. The success rate is computed as the percentage of frames where the intersection-over-union between the tracked bounding box and the groundtruth bounding box exceeds the given threshold. The success plots show that the ratios of successful frames at the thresholds vary from 0 to 1. In terms of success plots, we use area-under-curve (AUC) scores to summarize the trackers. All the numbers and plots are obtained using the standard OTB toolkit.

\begin{figure*}[t]
\centering
\includegraphics[width=55mm,height=47mm]{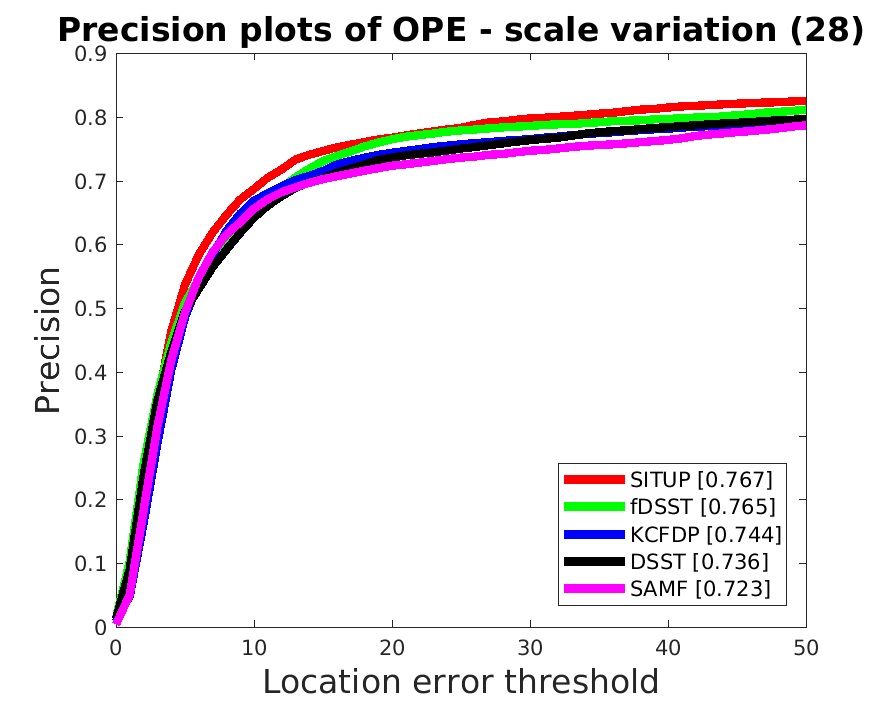}
\includegraphics[width=55mm,height=47mm]{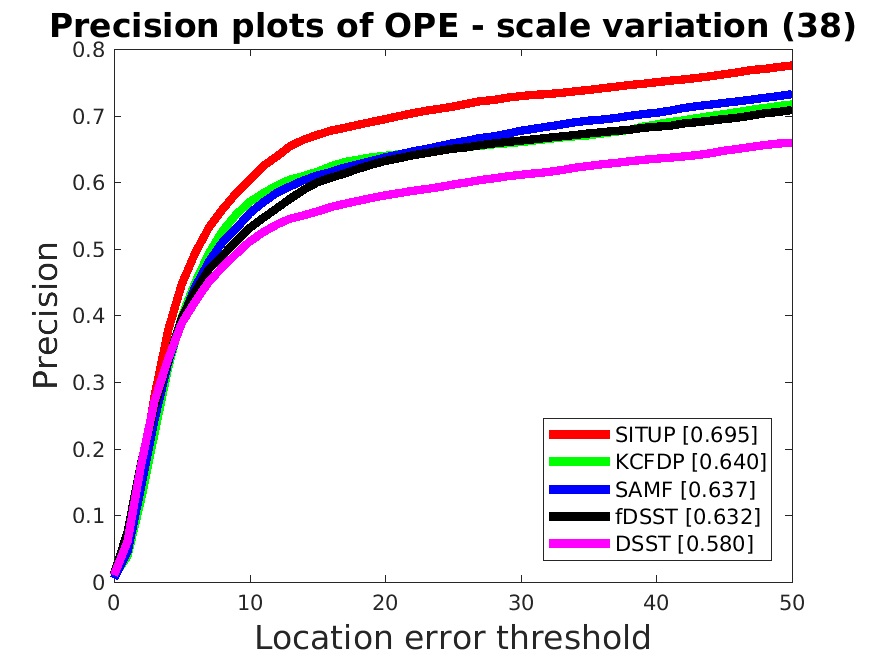}
\includegraphics[width=55mm,height=47mm]{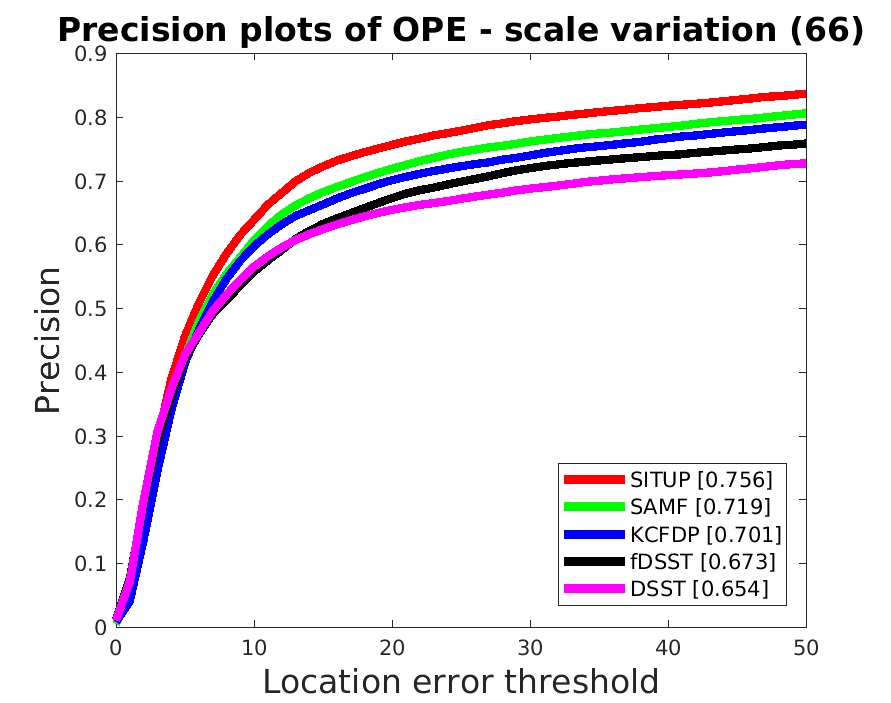}
\includegraphics[width=55mm,height=47mm]{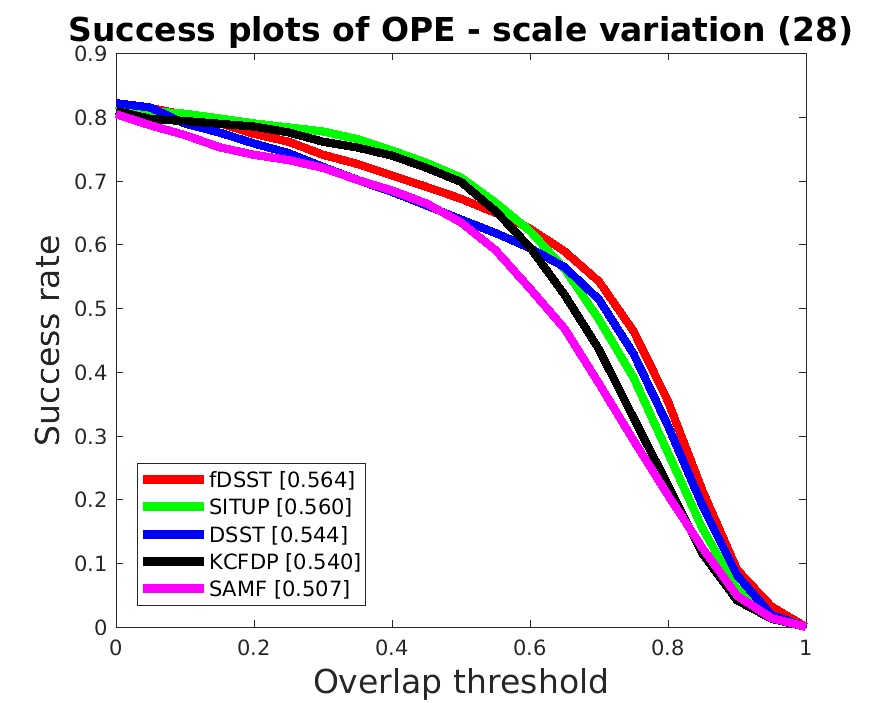}
\includegraphics[width=55mm,height=47mm]{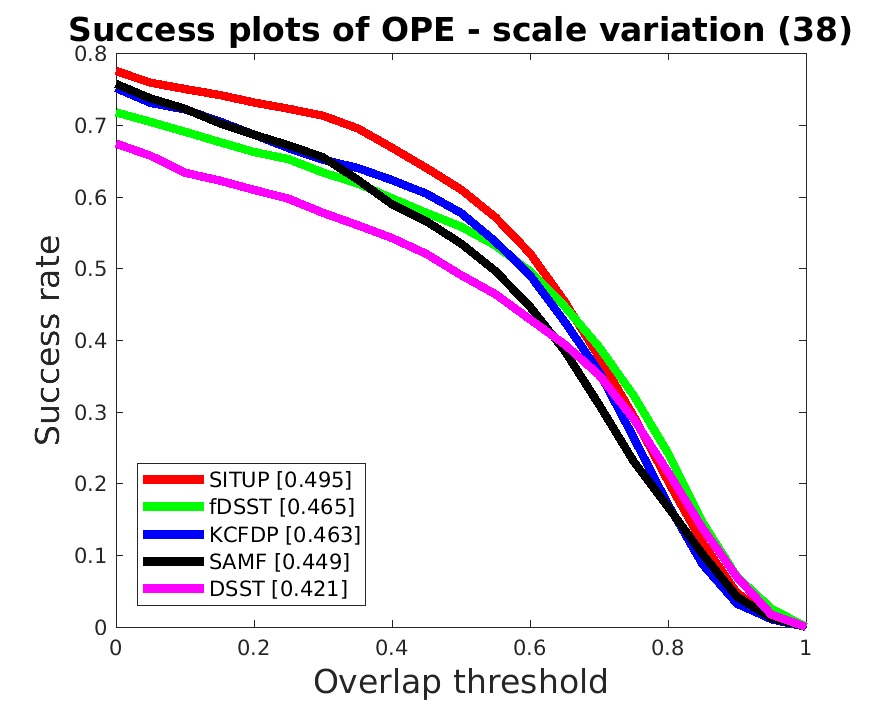}
\includegraphics[width=55mm,height=47mm]{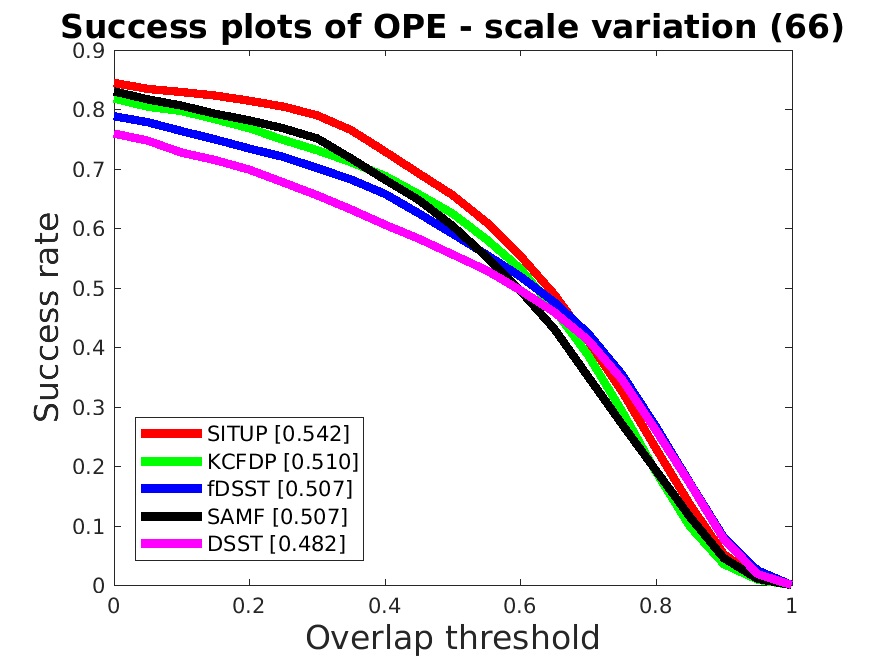}
\caption{The precision plots and success plots of our tracker SITUP and other scale adaptive variants of standard DCF based trackers on the scale variation attribute. From left to right, the plots correspond to the scale variation attribute of OTB2013, OTB50, OTB100. The top row are the precision plots and the bottom row are the success plots.}
\label{Fig5}
\end{figure*}                       

\begin{table}[t]
\caption{\upshape{Attributed-based comparisons with other scale adaptive variants of standard DCF based trackers on the scale variation (SV) attribute of OTB2013, OTB50 and OTB100 benchmark sequences. The first and the second best values are highlighted in bold and underlined.}}
\label{table}
    \begin{tabular}{cccccccc}
\hline
                       &                 & \multicolumn{2}{c}{\textbf{OTB2013}}        & \multicolumn{2}{c}{\textbf{OTB50}}          & \multicolumn{2}{c}{\textbf{OTB100}}         \\
\textbf{}              & \textbf{Method} & AUC                  & Prec.                & AUC                  & Prec.                & AUC                  & Prec.                \\ \hline
                       & DSST            & \multicolumn{1}{l}{0.544} & \multicolumn{1}{l}{0.736} & \multicolumn{1}{l}{0.421} & \multicolumn{1}{l}{0.580} & \multicolumn{1}{l}{0.482} & \multicolumn{1}{l}{0.654} \\
\multicolumn{1}{l}{}   & fDSST           & \multicolumn{1}{l}{\textbf{0.564}} & \multicolumn{1}{l}{\underline{0.765}} & \multicolumn{1}{l}{\underline{0.465}} & \multicolumn{1}{l}{0.632} & \multicolumn{1}{l}{0.507} & \multicolumn{1}{l}{0.673} \\
\multicolumn{1}{l}{SV} & KCFDP           & \multicolumn{1}{l}{0.540} & \multicolumn{1}{l}{0.744} & \multicolumn{1}{l}{0.463} & \multicolumn{1}{l}{\underline{0.640}} & \multicolumn{1}{l}{\underline{0.510}} & \multicolumn{1}{l}{0.701} \\
\multicolumn{1}{l}{}   & SAMF            & \multicolumn{1}{l}{0.507} & \multicolumn{1}{l}{0.723} & \multicolumn{1}{l}{0.449} & \multicolumn{1}{l}{0.637} & \multicolumn{1}{l}{0.507} & \multicolumn{1}{l}{\underline{0.719}} \\
\multicolumn{1}{l}{}   & SITUP            & \multicolumn{1}{l}{\underline{0.560}} & \multicolumn{1}{l}{\textbf{0.767}} & \multicolumn{1}{l}{\textbf{0.495}} & \multicolumn{1}{l}{\textbf{0.695}} & \multicolumn{1}{l}{\textbf{0.542}} & \multicolumn{1}{l}{\textbf{0.756}} \\ 
\hline
\end{tabular}
\end{table}
\subsection{DCF-based scale estimation comparison}
\label{333}

\subsubsection{Overall performance}

To compare the overall performance of SITUP to other DCF based trackers, we executed five trackers on the OTB benchmark datasets (DSST, fDSST, KCFDP, SAMF and our tracker SITUP). Table I summarizes the overall performance of the trackers in terms of AUC score and precision score for a threshold of 20 pixels (a typical parameter used in the literature). For completeness, the success plots and precision plots are given in Fig. 4. As shown in Table I and Fig. 4, the best results are obtained using SITUP. 

In comparison with fDSST, which obtains the second best performance on OTB2013 and OTB50, SITUP improves the AUC score by $1.0\%$, $1.2\%$ and $2.7\%$ and the precision score by $1.0\%$, $3.2\%$ and $6.0\%$ on OTB2013, OTB50, and OTB100 respectively.

In comparison with SAMF, which also employs a multi-resolution translation filter framework, SITUP improves the AUC score by $2.6\%$, $3.5\%$, and $3.3\%$ and the precision score by $2.7\%$, $5.2\%$ and $2.8\%$ on OTB2013, OTB50, and OTB100 respectively. It is worth mentioning that our approach achieves the largest performance improvement on OTB50, which contains the most difficult 50 sequences. As mentioned in Section \ref{SITUP}, the employed criterion APCE of our tracker is more robust than the maximum response value especially in challenging tracking scenarios, which is evidenced by the comparison between SITUP and SAMF.

\subsubsection{Attribute-based comparison}

Since SITUP is specifically designed for scale estimation, we compare SITUP with other scale adaptive DCF variants on the scale variation attribute of OTB2013, OTB50 and OTB100. OTB2013, OTB50 and OTB100 contain 28, 38 and 66 scale variation sequences, respectively, and can be used to evaluate the scale adaptability of our tracker. Table II summarizes the performance of the trackers in the AUC score and the precision score at a threshold of 20 pixels on the scale variation attribute. For completeness, the corresponding success plots and precision plots are shown in Fig. 5. 

\begin{figure*}[t]
\centering

\includegraphics[width=43mm,height=43mm]{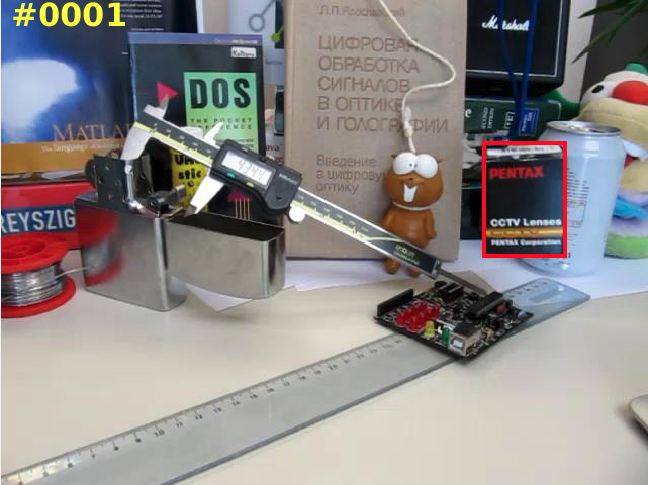}
\includegraphics[width=43mm,height=43mm]{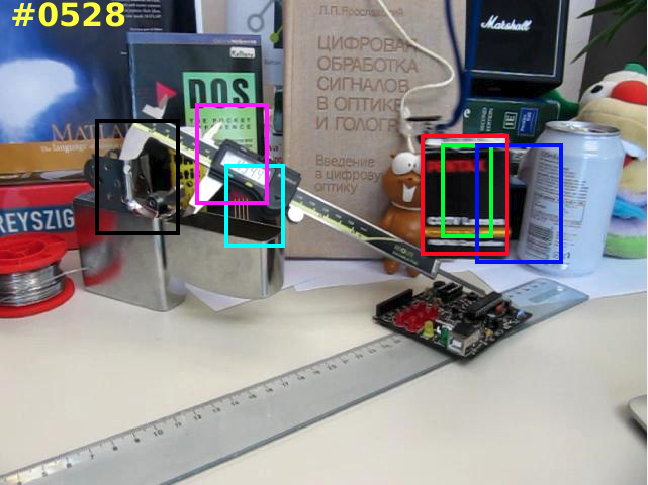}
\includegraphics[width=43mm,height=43mm]{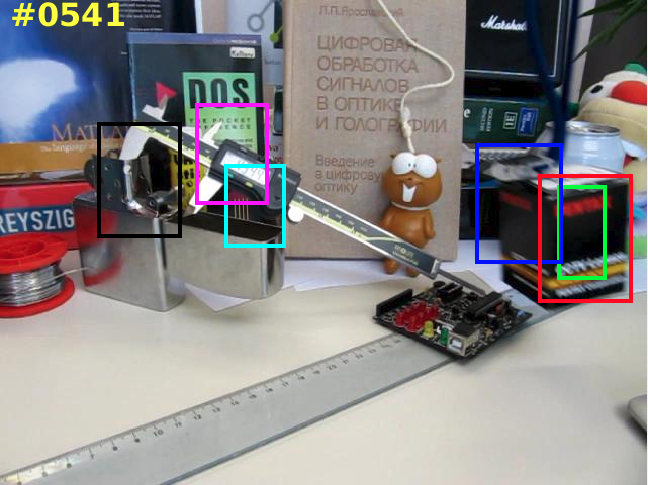}
\includegraphics[width=43mm,height=43mm]{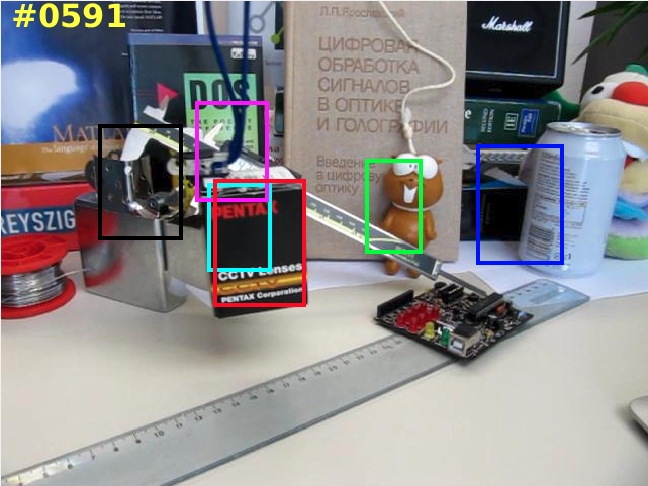}

\includegraphics[width=43mm,height=43mm]{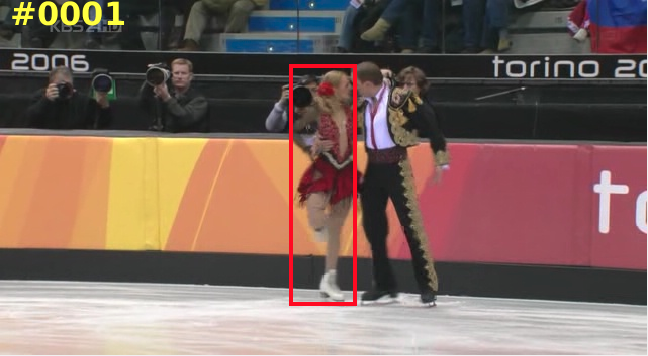}
\includegraphics[width=43mm,height=43mm]{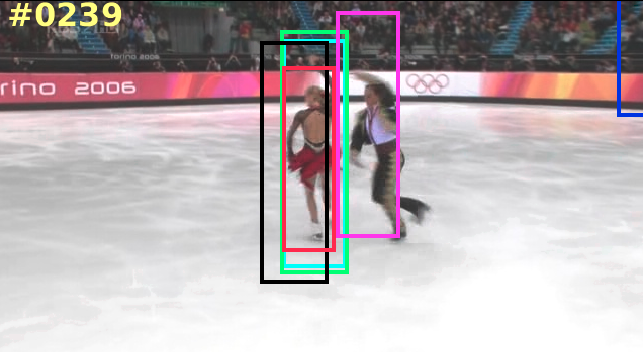}
\includegraphics[width=43mm,height=43mm]{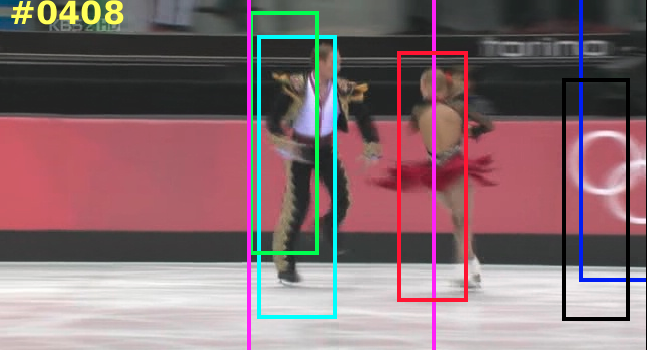}
\includegraphics[width=43mm,height=43mm]{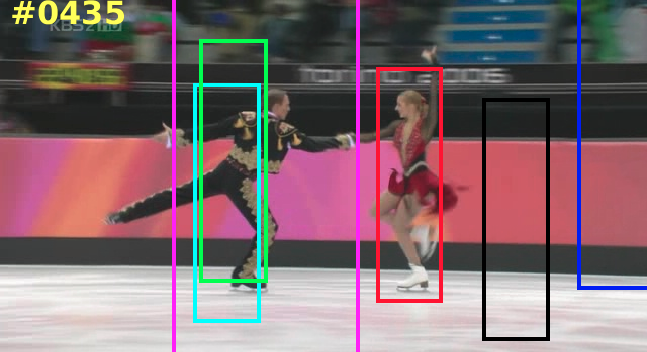}

\includegraphics[width=43mm,height=43mm]{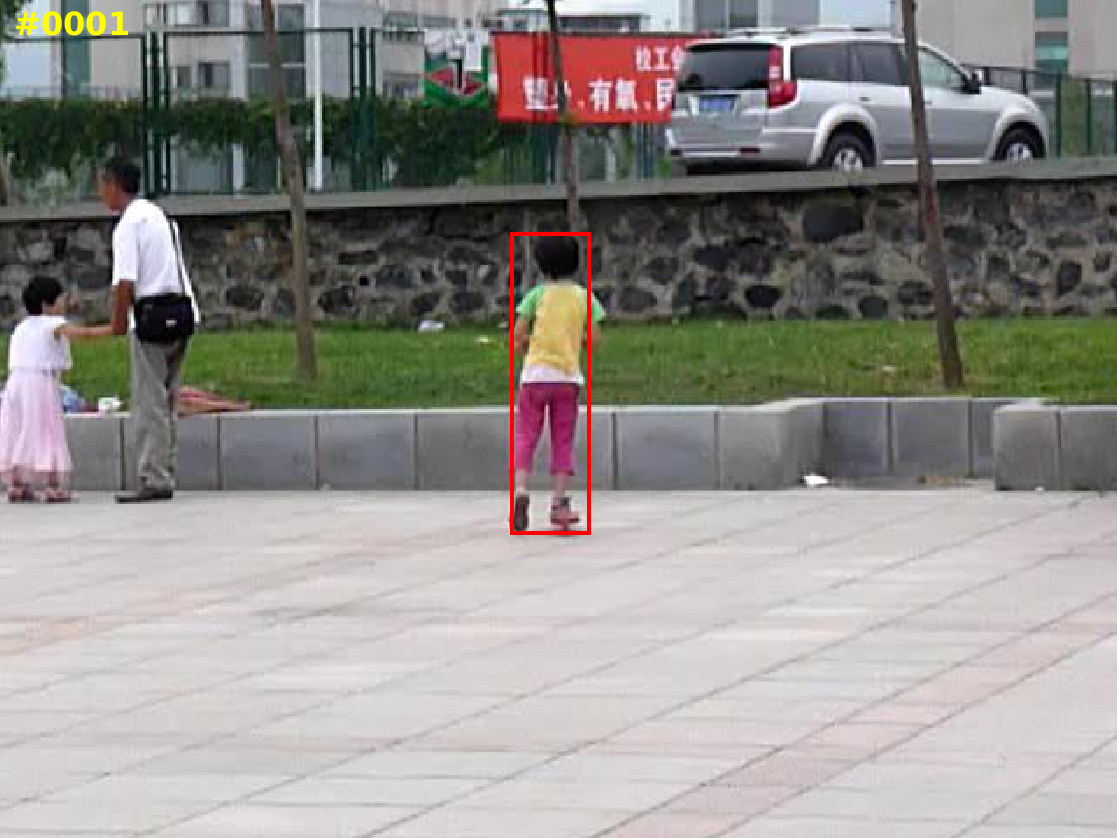}
\includegraphics[width=43mm,height=43mm]{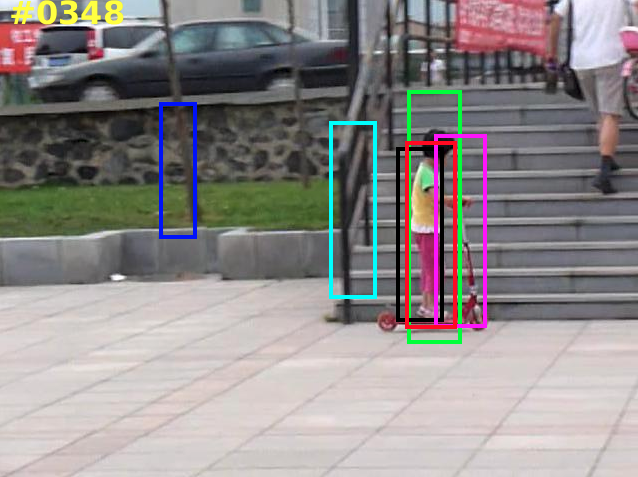}
\includegraphics[width=43mm,height=43mm]{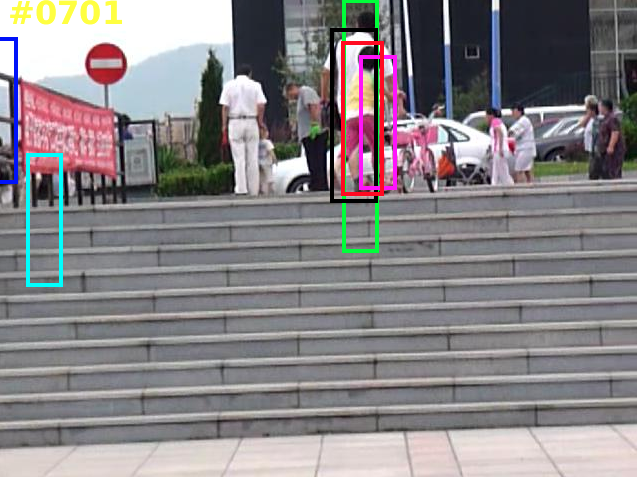}
\includegraphics[width=43mm,height=43mm]{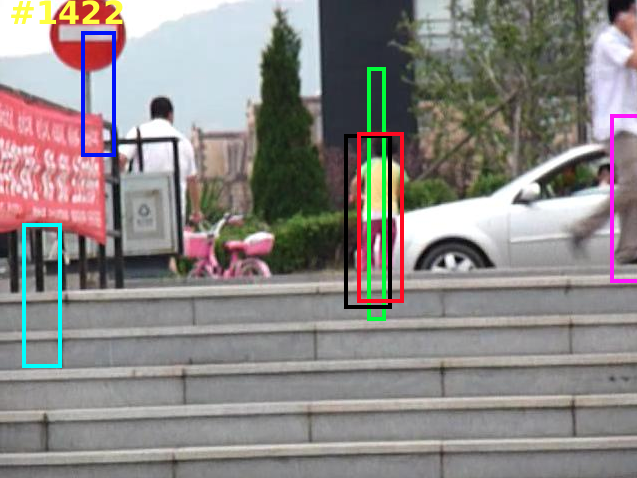}
\includegraphics[width=120mm,height=13mm]{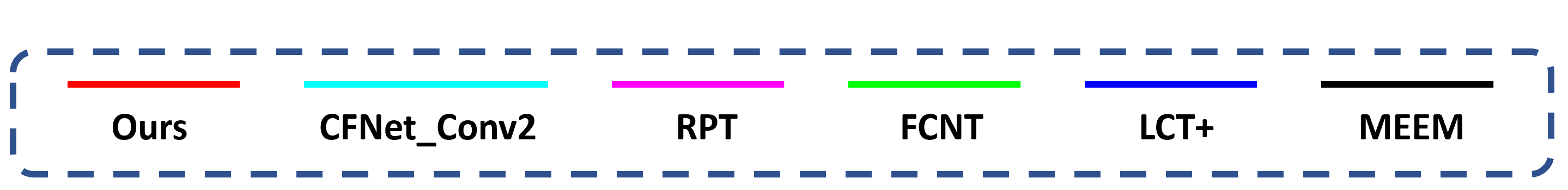}

\caption{Among 10 state-of-the-art trackers, CFNet$\_$Conv2, FCNT, LCT+, MEEM and RPT, which obtain top performance on OTB100, are selected to compare with our tracker. Example frames shown are from \textit{box} (top row), \textit{skating2$\_$1} (middle row) and \textit{girl2} (bottom row). The bounding boxes in the first column identify the target objects to be tracked in corresponding sequences, and the number on the upper-left corner of each image is the frame number of corresponding image. Our approach tracks the target object accurately in these challenging video sequences.}

\label{Fig6}
\end{figure*}

As shown in Table II and Fig. 5, our tracker achieves the best performance except for the AUC score on OTB2013, which is only $0.4\%$ lower than the best one obtained by fDSST. Comparing to fDSST, which obtains the second best overall performance on OTB2013 and OTB50, our approach improves the AUC score by $3.0\%$ and $3.5\%$ on more challenging OTB50 and OTB100, respectively. Also, our approach improves the precision score by $0.2\%$, $6.3\%$ and $8.3\%$ comparing to fDSST on OTB2013, OTB50 and OTB100, respectively. Comparing to KCFDP, which obtains the second best result on precision score of OTB50 and AUC score of OTB100, our approach improves the AUC score by $2.0\%$, $2.0\%$ and $3.2\%$ and the precision score by $2.3\%$, $5.5\%$ and $5.5\%$ on OTB2013, OTB50, and OTB100, respectively. The remarkable performance of our tracker shows that our method is superior in scale estimation comparing to other trackers despite the utter simplicity of our framework.

Comparing to SAMF, which also employs the multi-resolution translation filter framework, our tracker improves the AUC score by $5.3\%$, $4.6\%$ and $3.5\%$ and the precision score by $4.4\%$, $5.8\%$ and $3.7\%$ on OTB2013, OTB50 and OTB100, respectively. The comparison between our tracker and SAMF shows the effectiveness and robustness of the employed criterion APCE in the scale variation scenario.

\begin{table*}[t]
\caption{\upshape{Attributed-based comparison with representative state-of-the-art trackers on 11 attributes of OTB100 benchmark sequences. The first and the second best values are highlighted in bold and underlined. The 11 attributes in the OTB datasets are: illumination variation (IV), out-of-plane rotation (OPR), scale variation (SV), occlusion (OCC), deformation (DEF), motion blur (MB), fast motion (FM), in-plane rotation (IPR), out-of-view (OV),  background clutter (BC), and low resolution (LR).}}
\resizebox{\textwidth}{24mm}{
\begin{tabular}{ccccccccccccccccccccccccc}
\hline
             & \multicolumn{2}{c}{Overall}     & \multicolumn{2}{c}{SV}          & \multicolumn{2}{c}{MB}          & \multicolumn{2}{c}{IPR}          & \multicolumn{2}{c}{OPR}          & \multicolumn{2}{c}{FM}          & \multicolumn{2}{c}{DEF}         & \multicolumn{2}{c}{LR}          & \multicolumn{2}{c}{OV}          & \multicolumn{2}{c}{BC}          & \multicolumn{2}{c}{IV}          & \multicolumn{2}{c}{OCC}         \\
Method       & AUC            & Prec           & AUC            & Prec           & AUC            & Prec           & AUC            & Prec           & AUC            & Prec           & AUC            & Prec           & AUC            & Prec           & AUC            & Prec           & AUC            & Prec           & AUC            & Prec           & AUC            & Prec           & AUC            & Prec           \\ \hline
CFNet\_Conv2 & {\ul{ 0.568}}    & 0.748          & \textbf{0.544} & 0.720          & {\ul 0.567}    & 0.687          & \textbf{0.573} & 0.771          & {\ul 0.537}    & 0.711          & \textbf{0.553} & 0.695          & 0.470          & 0.638          & \textbf{0.590} & \textbf{0.787} & 0.414          & 0.533          & 0.538          & 0.715          & 0.526          & 0.677          & {\ul 0.540}    & 0.720          \\
FCNT         & 0.428          & 0.780          & 0.398          & \textbf{0.757} & 0.417          & 0.691          & 0.455          & \textbf{0.808} & 0.427          & {\ul 0.772}    & 0.422          & {\ul 0.707}    & 0.380          & \textbf{0.750} & 0.336          & {\ul 0.739}    & 0.348          & 0.596          & 0.430          & {\ul 0.779}    & 0.460          & \textbf{0.795} & 0.415          & 0.750          \\
LCT+         & 0.562          & 0.762          & 0.497          & 0.691          & 0.532          & 0.673          & {\ul 0.562}    & 0.786          & 0.530          & 0.717          & 0.522          & 0.667          & {\ul 0.490}    & 0.675          & 0.330          & 0.490          & 0.452          & 0.592          & 0.542          & 0.725          & \textbf{0.553} & 0.732          & 0.536          & 0.742          \\
TLD          & 0.595          & 0.426          & 0.396          & 0.575          & 0.434          & 0.546          & 0.436          & 0.617          & 0.391          & 0.565          & 0.424          & 0.553          & 0.345          & 0.490          & 0.372          & 0.552          & 0.351          & 0.474          & 0.346          & 0.446          & 0.397          & 0.535          & 0.378          & 0.552          \\
MEEM         & 0.530          & {\ul 0.781}    & 0.479          & 0.744          & 0.545          & {\ul 0.722}    & 0.534          & {\ul 0.798}    & 0.521          & \textbf{0.782} & 0.525          & \textbf{0.728} & 0.484          & {\ul 0.746}    & 0.355          & 0.605          & {\ul 0.488}    & \textbf{0.685} & 0.514          & 0.737          & 0.504          & 0.725          & 0.526          & {\ul 0.775}    \\
TGPR         & 0.458          & 0.643          & 0.401          & 0.596          & 0.409          & 0.508          & 0.463          & 0.665          & 0.440          & 0.616          & 0.398          & 0.507          & 0.449          & 0.623          & 0.378          & 0.629          & 0.373          & 0.493          & 0.440          & 0.610          & 0.461          & 0.639          & 0.454          & 0.646          \\
STRUCK       & 0.461          & 0.638          & 0.414          & 0.609          & 0.459          & 0.592          & 0.459          & 0.640          & 0.415          & 0.566          & 0.455          & 0.609          & 0.379          & 0.521          & 0.347          & 0.628          & 0.374          & 0.487          & 0.425          & 0.544          & 0.400          & 0.524          & 0.425          & 0.588          \\
RPT          & 0.536          & 0.756          & 0.489          & 0.718          & 0.510          & 0.711          & 0.529          & 0.749          & 0.510          & 0.714          & 0.525          & 0.706          & 0.488          & 0.701          & 0.358          & 0.575          & 0.475          & 0.609          & \textbf{0.572} & \textbf{0.793} & 0.526          & {\ul 0.791}    & 0.491          & 0.697          \\
KCF          & 0.477          & 0.696          & 0.405          & 0.644          & 0.456          & 0.618          & 0.476          & 0.707          & 0.449          & 0.643          & 0.448          & 0.620          & 0.433          & 0.612          & 0.307          & 0.546          & 0.393          & 0.501          & 0.492          & 0.704          & 0.464          & 0.703          & 0.465          & 0.689          \\
CSK          & 0.382          & 0.518          & 0.332          & 0.465          & 0.314          & 0.373          & 0.387          & 0.524          & 0.359          & 0.476          & 0.326          & 0.400          & 0.333          & 0.443          & 0.263          & 0.389          & 0.250          & 0.276          & 0.401          & 0.560          & 0.346          & 0.453          & 0.351          & 0.474          \\
SITUP       & \textbf{0.576} & \textbf{0.782} & {\ul 0.542}    & {\ul 0.756}    & \textbf{0.579} & \textbf{0.725} & 0.545          & 0.744          & \textbf{0.552} & 0.739          & {\ul 0.537}    & 0.701          & \textbf{0.513} & 0.708          & {\ul 0.464}    & 0.709          & \textbf{0.520} & {\ul 0.669}    & {\ul 0.564}    & 0.769          & {\ul 0.542}    & 0.727          & \textbf{0.572} & \textbf{0.780} \\ \hline
\end{tabular}}
\end{table*}

\subsection{State-of-the-art comparison}
\label{444}

To demonstrate the superior performance of our approach, we compare our tracker to 10 representative state-of-the-art trackers. These trackers can be broadly categorized as follows:
\begin{itemize}
\item Deep learning based trackers, such as FCNT \cite{FCNT}, which employ deep features in correlation filter framework, and CFNet$\_$Conv2 \cite{cfnet}. Such trackers train the correlation filter end-to-end with ILSVRC15-VID containing almost 4500 videos with a total of more than one million annotated frames.
\item Trackers which are designed for long-term tracking by employing both short-term tracker and online classifier, including LCT+ \cite{LCT+} and TLD \cite{TLD}.
\item Representative trackers that employ single or multiple online classifiers, including MEEM \cite{MEEM}, TGPR \cite{TGPR}, and Struck \cite{STRUCK} methods.
\item Representative baseline trackers of correlation filter trackers KCF \cite{KCF} and CSK \cite{CSK}.\
\item Representative part-based tracking (RPT) methods \cite{RPT} which exploit reliable patches.
\end{itemize}

For the sake of simplicity, all tracking algorithms in this subsection are only evaluated on the OTB100 dataset since OTB2013 and OTB50 are contained in the OTB100 dataset. The overall and attribute-based AUC scores and precision scores of all trackers are reported in Table III. The qualitative comparison is presented in Fig. 6 and the plots for the overall performance of all trackers on OTB100 are presented in Fig. 7.

\begin{figure}[t]
\centering
\includegraphics[width=43mm,height=50mm]{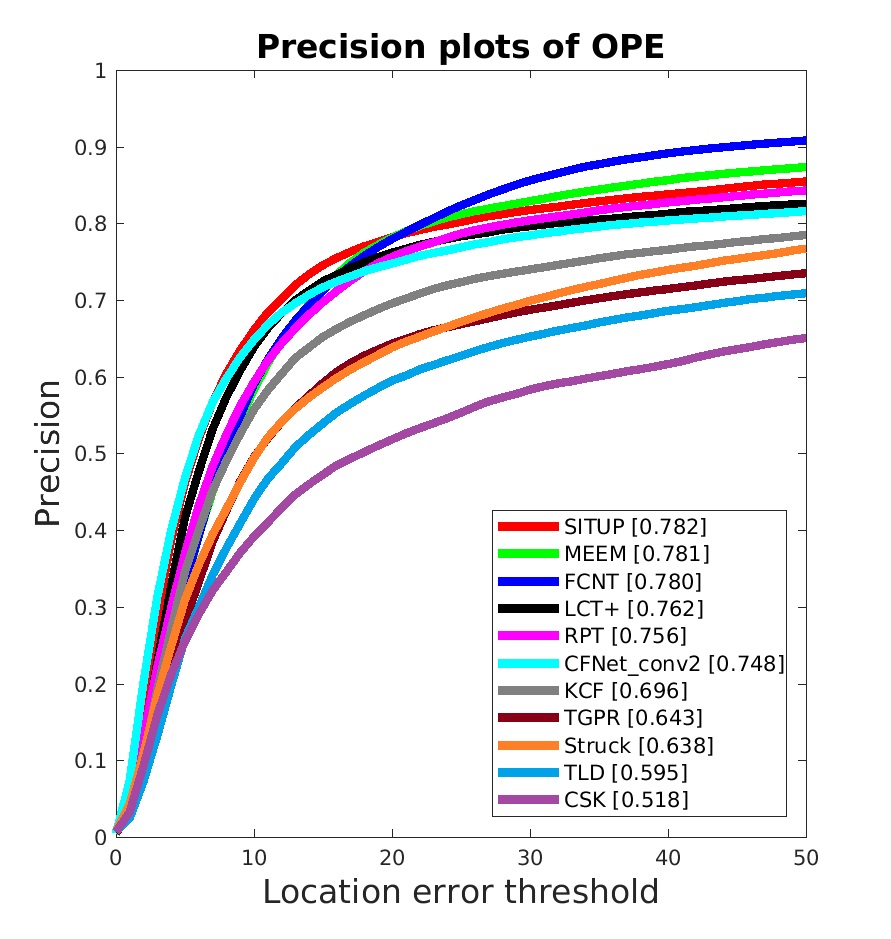}
\includegraphics[width=43mm,height=50mm]{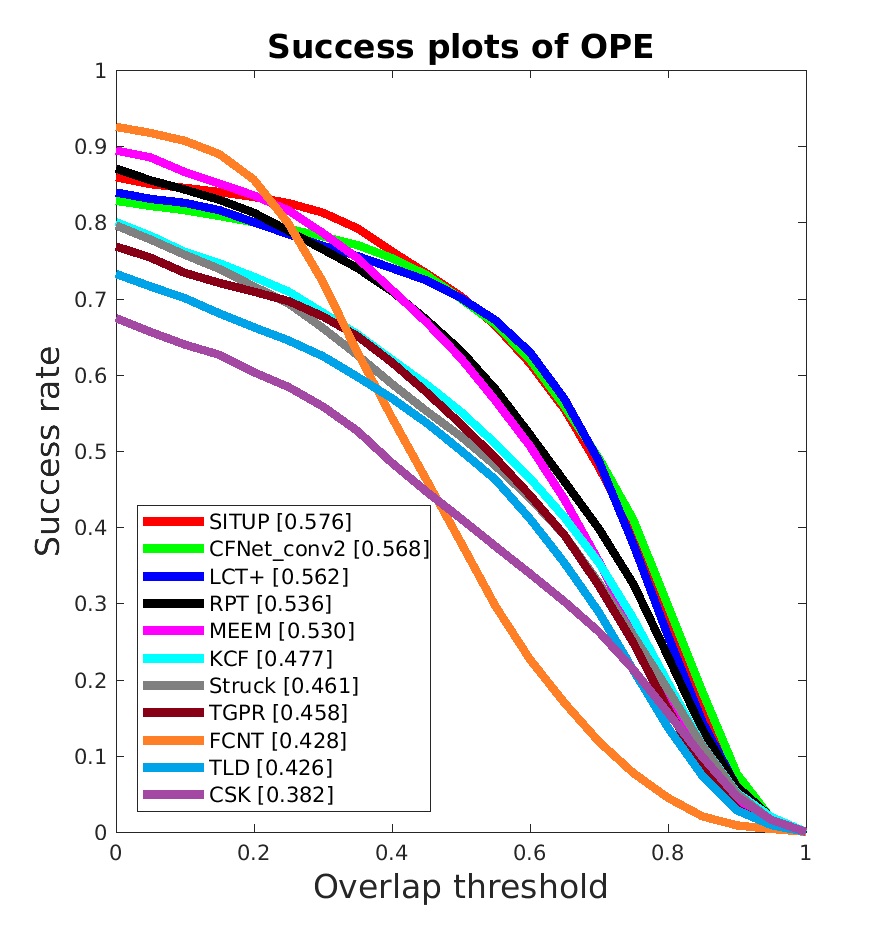}
\caption{The precision plots and success plots of our tracker and other state-of-the-art trackers for the overall performance on the OTB100 benchmark dataset.}
\label{Fig5}
\end{figure}

\subsubsection{Overall performance}

As shown in Table III, our tracker achieves the best performance in terms of the AUC score and the precision score despite the utter simplicity of its framework and its real-time running speed on a CPU. Comparing to CFNet$\_$Conv2, which achieves the second best AUC score, our tracker improves the AUC score by $0.8\%$ and the precision score by $3.4\%$. Our tracker improves the AUC score by $4.6\%$ compared to MEEM, which achieves the second best precision score. Although MEEM and CFNet$\_$Conv2 obtain favorable performance compared to our tracker, the speed of MEEM is only about 20 fps, and CFNet$\_$Conv2 needs a high-end GPU for its implementation.

\subsubsection{Attribute-based comparison}

For the scale variation attribute, SITUP obtains the second best AUC score and the second best precision score, which are only $0.2\%$ and $0.1\%$ lower than the best ones, which are obtained by the deep learning based methods CFNet$\_$Conv2 and FCNT, respectively. It is worth mentioning that SITUP obtains favorable performance in 10 of 11 attributes with the exception being the low resolution challenge. These promising results suggest that the effectiveness of our tracking framework and the robustness of our employed APCE measure when facing different scenarios. In terms of the low resolution challenge, our performance is $12.6\%$ and $7.8\%$ lower than the best ones obtained by CFNet$\_$Conv2 on AUC score and precision score, respectively. It is likely that the performance gap mainly comes from the features employed. The features we employed are HoG, CN and grayscale pixel values, which mainly focus on the texture information and color information. However, the deep features employed in CFNet$\_$Conv2 contain semantic information that is more discriminative when low resolution is encountered.

\section{Conclusions}
\label{conclusion}

In this paper, an effective scale estimation approach is proposed in SITUP to address the problem of fixed template size in standard discriminative correlation filter (DCF) based trackers by incorporating a novel criterion named average peak-to-correlation energy (APCE) into the multi-resolution translation filter framework. Our generic scale estimation approach can be incorporated into any DCF based tracker.

Extensive experiments are conducted on the OTB datasets OTB2013, OTB50 and OTB100. The comparison between SITUP and other scale adaptive variants of standard DCF based trackers clearly demonstrates the effectiveness of our scale searching strategy and the robustness of the employed criterion APCE. Also, the comparison between SITUP and 10 state-of-the-art trackers demonstrates superior performance of our tracker. Although our tracker is specifically designed for the scale variation challenge, the promising results on other challenging attributes again evidences the robustness of our employed APCE measure when facing different scenarios.


%


\ifCLASSOPTIONcaptionsoff
  \newpage
\fi




\begin{thebibliography}{99}
\bibitem{video surveillance}
A. Hampapur et al., ``Smart video surveillance: exploring the concept of multiscale spatiotemporal tracking," \textit{IEEE Signal Processing Magazine}, vol.22, no.2, pp. 38-51, 2005.
\bibitem{video compression}
L. Itti, ``Automatic foveation for video compression using a neurobiological model of visual attention," {\em IEEE Transactions on Image Processing}, vol. 13, no.2, pp. 1304-1318, 2004.
\bibitem{ray2002tracking}
N. Ray, S. T. Acton, and K. Ley, ``Tracking leukocytes in vivo with shape and size constrained active contours," 
{\em IEEE Transactions on Medical Imaging}, 
vol. 21, no. 10, pp. 1222-1235, 2002.
\bibitem{cui2006monte}
J. Cui, S. T. Acton, and Z. Lin, ``A Monte Carlo approach to rolling leukocyte tracking in vivo," \textit{Medical Image Analysis}, vol. 10, no. 4, pp.598-610, 2006.
\bibitem{liu2012grid}
X. Liu, Z. Lin, and S. T. Acton, ``A grid-based {B}ayesian approach to robust visual tracking," \textit{Digital Signal Processing}, vol. 22, no. 1, pp. 54-65, 2012.
\bibitem{mukherjee2004level}
D.P. Mukherjee, N. Ray, and S. T. Acton, ``Level set analysis for leukocyte detection and tracking," \textit{IEEE Transactions on Image processing}, vol. 13, no. 4, pp.562-572, 2004.
\bibitem{IVT}
D. A. Ross, J. Lim, R.-S. Lin, and M.-H. Yang, ``Incremental learning for robust visual tracking," \textit{International Journal of Computer Vision}, vol. 77, no. 1-3, pp. 125-141, 2008.
\bibitem{l1tracking}
M. Xue and H. Ling, `` Robust visual tracking using $\ell_1$ minimization," in \textit{International Conference on Computer Vision (ICCV)}, 2009.
\bibitem{HOG}
N. Dalal, and B. Triggs, ``Histograms of oriented gradients for human detection," in \textit{Computer Vision and Pattern Recognition (CVPR)}, 2005.
\bibitem{color}
H. Possegger, T. Mauthner, and H. Bischof, ``In defense of color-based model-free tracking," in \textit{Computer Vision and Pattern Recognition (CVPR)}, 2015.
\bibitem{TLD}
Z. Kalal, K. Mikolajczyk, and J. Matas, ``Tracking-leaning-detection," \textit{IEEE Transactions on Pattern Analysis and Machine Intelligence,} vol. 34, no. 7, pp. 1409-1422, 2012.
\bibitem{STRUCK}
S. Hare, S. Golodetz, A. Saffari, V. Vineet, M.-M Cheng, S. L. Hicks, and P. H. Torr, ``Struck: Structured output tracking with kernels," \textit{IEEE Transactions on Pattern Analysis and Machine Intelligence,} vol. 38, no. 10, pp. 2096-2109, 2016.
\bibitem{MEEM}
J. Zhang, S. Ma, and S. Sclaroff, ``MEEM: Robust tracking via multiple experts using entropy minimization," in \textit{European Conference on Computer Vision,} 2014.
\bibitem{MOSSE}
D. S. Bolme, J. R. Beveridge, B. A. Draper, and Y. M. Lui, ``Visual object tracking using adaptive correlation filters," in \textit{Computer Vision and Pattern Recognition (CVPR)}, 2010.
\bibitem{KCF}
J. F. Henriques, R. Caseiro, P. Martins, and J. Batista, ``High-speed tracking with kernelized correlation filters," \textit{IEEE Transactions on Pattern Analysis and Machine Intelligence,} vol. 37, no. 3, pp. 583-596, 2015.
\bibitem{DSST&fDSST}
M. Danelljan, G. H{\"a}ger, F. S. Khan, and M. Felsberg, ``Discriminative scale space tracking," \textit{IEEE Transactions on Pattern Analysis and Machine Intelligence}, vol. 39, no. 8, pp. 1561-1575, 2017.
\bibitem{zhang2015robust}
M. Zhang, J. Xing, J. Gao, and W. Hu, ``Robust visual tracking using joint scale-spatial correlation filters,” in \textit{International Conference on Image Processing (ICIP)}, 2015.
\bibitem{kcfdp}
D. Huang, L. Luo, M. Wen, Z. Chen, and C. Zhang, ``Enable scale and aspect ratio adaptability in visual tracking with detection proposals," in \textit{British Machine Vision Conference (BMVC)}, 2015.
\bibitem{skcf}
A. S. Montero, J. Lang, and R. Laganiere, ``Scalable kernel correlation filter with sparse feature integration,” in \textit{International Conference on Computer Vision Workshop (ICCVW)}, 2015.
\bibitem{OTB50}
Y. Wu, J. Lim, and M.-H. Yang, ``Online object tracking: A benchmark," in \textit{Computer Vision and Pattern Recognition (CVPR)}, 2013.
\bibitem{OTB100}
Y. Wu, J. Lim, and M.-H. Yang, ``Object tracking benchmark," \textit{IEEE Transactions on Pattern Analysis and Machine Intelligence}, vol. 37, no. 9, pp. 1834-1848, 2015.
\bibitem{VOT2014}
M. Kristan et al., ``The visual object tracking VOT2014 challenge results," \textit{European Conference on Computer Vision (ECCV)}, 2014.
\bibitem{prokaj2014persistent}
J. Prokaj and G. Medioni, ``Persistent tracking for wide area aerial surveillance," in \textit{Computer Vision and Pattern Recognition (CVPR)}, 2014.
\bibitem{danelljan2014low}
M. Danelljan, F. S. Khan, M. Felsberg, K. Granstr{\"o}m, F. Heintz, P. Rudol, M. Wzorek, J. Kvarnstr{\"o}m, and P. Doherty, ``A low-level active vision framework for collaborative unmanned aircraft systems," in \textit{European Conference on Computer Vision Workshop (ECCVW)}, 2014.
\bibitem{geiger20143d}
A. Geiger, M. Lauer, C. Wojek, C. Stiller, and R. Urtasun, ``3d traffic scene understanding from movable platforms," \textit{IEEE Transactions on
Pattern Analysis and Machine Intelligence}, vol. 36, no. 5, pp. 1012-1025, 2014.
\bibitem{CSK}
J. F. Henriques, R. Caseiro, P. Martins, and J. Batista, ``Exploiting the circulant structure of tracking-by-detection with kernels," in \textit{European
Conference on Computer Vision (ECCV)}, 2012.
\bibitem{CN}
M. Danelljan, F. Shahbaz Khan, M. Felsberg, and J. Van de Weijer, ``Adaptive color attributes for real-time visual tracking," in \textit{Conference
on Computer Vision and Pattern Recognition (CVPR)}, 2014.
\bibitem{edgebox}
C. L. Zitnick and P. Doll{\'a}r, ``Edge boxes: Locating object proposals from edges," in \textit{European Conference on Computer Vision (ECCV)}, 2014.
\bibitem{gray2006toeplitz}
R. M. Gray et al., ``Toeplitz and circulant matrices: A review," \textit{Foundations and Trends in Communications and Information Theory}, vol. 2, no. 3, pp. 155-239, 2006.
\bibitem{khan2013coloring}
F. S. Khan, R. M. Anwer, J. Van De Weijer, A. D. Bagdanov, A. M. Lopez, and M. Felsberg, ``Coloring action recognition in still images," \textit{International Journal of Computer Vision}, vol. 105, no. 3, pp. 205-221, 2013.
\bibitem{khan2012color}
F. S. Khan, R. M. Anwer, J. Van de Weijer, A. D. Bagdanov, M. Vanrell, and A. M. Lopez, ``Color attributes for object detection," in \textit{Computer
Vision and Pattern Recognition (CVPR)}, 2012.
\bibitem{khan2012modulating}
F. S. Khan, J. Van de Weijer, and M. Vanrell, ``Modulating shape features by color attention for object recognition," \textit{International Journal
of Computer Vision (ICCV)}, vol. 98, no. 1, pp. 49-64, 2012.
\bibitem{staple}
L. Bertinetto, J. Valmadre, S. Golodetz, O. Miksik, and P. H. Torr, ``Staple: Complementary learners for real-time tracking," in \textit{Computer Vision and Pattern Recognition (CVPR)}, 2016.
\bibitem{siena2016detecting}
S. Siena and B. V. Kumar, ``Detecting occlusion from color information to improve visual tracking," in \textit{IEEE International Conference on Acoustics, Speech and Signal Processing (ICASSP)}, 2016.
\bibitem{felzenszwalb2010object}
P. F. Felzenszwalb, R. B. Girshick, D. McAllester, and D. Ramanan, ``Object detection with discriminatively trained part-based models," \textit{IEEE Transactions on Pattern Analysis and Machine Intelligence}, vol. 32, no. 9, pp. 1627-1645, 2010.
\bibitem{LMCF}
M. Wang, Y. Liu, and Z. Huang, ``Large margin object tracking with circulant feature maps," in \textit{Computer Vision and Pattern Recognition (CVPR)}, 2017.
\bibitem{FCNT}
L. Wang, W. Ouyang, X. Wang, and H. Lu, ``Visual tracking with fully convolutional networks," in \textit{International Conference on Computer Vision (ICCV)}, 2015.
\bibitem{cfnet}
J. Valmadre, L. Bertinetto, J. Henriques, A. Vedaldi, and P. H. Torr, ``End-to-end representation learning for correlation filter based tracking," in \textit{Computer Vision and Pattern Recognition (CVPR)}, 2017.
\bibitem{LCT+}
C. Ma, J.-B. Huang, X. Yang, and M.-H. Yang, ``Adaptive correlation filters with long-term and short-term memory for object tracking," \textit{International Journal of Computer Vision}, pp. 1-26, 2018.
\bibitem{TGPR}
J. Gao, H. Ling, W. Hu, and J. Xing, ``Transfer learning based visual tracking with gaussian processes regression," in \textit{European Conference
on Computer Vision (ECCV)}, 2014.
\bibitem{RPT}
Y. Li, J. Zhu, and S. C. Hoi, ``Reliable patch trackers: Robust visual tracking by exploiting reliable patches," in \textit{Computer Vision and Pattern Recognition (CVPR)}, 2015.
\bibitem{SAMF}
Y. Li and J. Zhu, ``A scale adaptive kernel correlation filter tracker with feature integration," in \textit{European Conference on Computer Vision (ECCV)}, 2014.
\bibitem{DSST}
M. Danelljan, G. H{\"a}ger, F. Khan, and M. Felsberg, ``Accurate scale estimation for robust visual tracking," in \textit{British Machine Vision Conference (BMVC)}, 2014.
\bibitem{gen1}
S. Zhou, R. Chellappa, and B. Moghaddam, ``Visual tracking and recognition using appearance-adaptive models in particle filters," \textit{IEEE Transactions on Image Processing}, vol. 13, no. 11, pp. 1491-1506, 2004.
\bibitem{gen2}
A. D. Jepson, D. J. Fleet, and T. F. EI-Maraghi, ``Robust online appearance models for visual tracking," in \textit{Computer Vision and Pattern Recognition (CVPR)}, 2003.
\end{thebibliography}
%

\end{document}